\documentclass[3p,twocolumn,coverpage,sort&compress]{elsarticle}
\usepackage{amsmath} 
\usepackage[hyperindex,breaklinks]{hyperref} 
\usepackage{atlasphysics}
\usepackage{booktabs}
\usepackage{lineno}
\usepackage{multirow}
\usepackage{bigstrut}
\usepackage{subfigure}
\newcommand\jet{\mathrm{jet}}

\newcommand\mbb{\ensuremath{m_{\bbbar}}}

\def\ptZ{\ensuremath{p_{\mathrm{T}}^{Z}}} 
\def\ptW{\ensuremath{p_{\mathrm{T}}^{W}}} 
\def\ptV{\ensuremath{p_{\mathrm{T}}^{V}}} 
\def\MPT{\ensuremath{p_{\mathrm{T}}^{\mathrm{miss}}}} 

\newcommand\whtolnubb{$WH \to \ell\nu b\bar{b}$\ }
\newcommand\zhtollbb{$ZH \to \ell^+\ell^- b\bar{b}$\ }
\newcommand\zhtovvbb{$ZH \to \nu\bar{\nu} b\bar{b}$\ }
\DeclareGraphicsExtensions{.pdf}
\usepackage{preprintcover}  

\PreprintCoverPaperTitle{Search for the Standard Model Higgs boson produced in association with a vector boson and decaying to a $b$-quark pair with the ATLAS detector}  

\PreprintIdNumber{CERN-PH-EP-2012-138}  

\PreprintCoverAbstract{This Letter presents the results of a direct search with the ATLAS
detector at the LHC for a Standard Model Higgs boson of mass $110 \le
\mH \le 130$~\GeV\ produced in association with a $W$ or $Z$ boson and
decaying to $\bbbar$. Three decay channels are considered:
$ZH\rightarrow \ell^+\ell^-b\bar{b}$, $WH\rightarrow \ell\nu b\bar{b}$
and $ZH\rightarrow \nu\bar{\nu}b\bar{b}$, where $\ell$ corresponds to
an electron or a muon.  No evidence for Higgs boson production is
observed in a dataset of $7$~\TeV\ $pp$ collisions corresponding to
$4.7$~\ifb\ of integrated luminosity collected by ATLAS in
2011. Exclusion limits on Higgs boson production, at the $95\%$
confidence level, of $2.5$ to $5.5$ times the Standard Model cross
section are obtained in the mass range $110$--$130\GeV$. The expected
exclusion limits range between $2.5$ and $4.9$ for the same mass
interval.}  

\PreprintJournalName{Phys. Lett. B.}  

\def\papertitle{Search for the Standard Model Higgs boson produced in association with a vector boson and decaying to a $b$-quark pair with the ATLAS detector}

\def\theabstract{%
 This Letter presents the results of a direct search with the ATLAS
detector at the LHC for a Standard Model Higgs boson of mass $110 \le
\mH \le 130$~\GeV\ produced in association with a $W$ or $Z$ boson and
decaying to $\bbbar$. Three decay channels are considered:
$ZH\rightarrow \ell^+\ell^-b\bar{b}$, $WH\rightarrow \ell\nu b\bar{b}$
and $ZH\rightarrow \nu\bar{\nu}b\bar{b}$, where $\ell$ corresponds to
an electron or a muon.  No evidence for Higgs boson production is
observed in a dataset of $7$~\TeV\ $pp$ collisions corresponding to
$4.7$~\ifb\ of integrated luminosity collected by ATLAS in
2011. Exclusion limits on Higgs boson production, at the $95\%$
confidence level, of $2.5$ to $5.5$ times the Standard Model cross
section are obtained in the mass range $110$--$130\GeV$. The expected
exclusion limits range between $2.5$ and $4.9$ for the same mass
interval.}
\begin{document}

  \begin{frontmatter}
    \title{\papertitle}
    \author{The ATLAS Collaboration}
    \address{}
    \begin{abstract} 
      \theabstract
    \end{abstract}
  \end{frontmatter}

  \section{Introduction}
The search for the Standard Model (SM) Higgs
boson~\cite{Englert:1964et,Higgs:1964pj,Guralnik:1964eu} is one of the
most important endeavours of the Large Hadron Collider (LHC). The
results of searches in various channels using data corresponding to an
integrated luminosity of up to $4.9\, \ifb$ have been reported
recently by both the ATLAS and CMS
collaborations~\cite{atlas:2012comb,cms:2012comb}.  The Higgs boson
has been excluded at the $95\%$ confidence level below $114.4\GeV$ by
the LEP experiments~\cite{Barate:2003sz}, in the regions
100--$106\gev$ and 147--$179\gev$ at the Tevatron $p\bar{p}$
collider~\cite{tevatron:higgs_comb2011}, and in the regions
$112.5$--$115.5\GeV$ and $127$--$600\GeV$ by the LHC experiments.
This Letter reports on a search for the SM Higgs boson performed for
the $H \to \bbbar$ decay mode, over the mass range $110$-$130\GeV$
where this decay mode dominates.

Due to the large backgrounds present in the dominant production
process $gg \to H \to \bbbar$, the analysis reported here is
restricted to Higgs boson production in association with a vector
boson, $WH$ and
$ZH$~\cite{Han:1991ia,Brein:2003wg,Ciccolini:2003jy,Denner:2011id,Ferrera:2011bk},
where the vector boson provides an additional final state signature,
allowing for significant background suppression.  An additional 
handle against the backgrounds is provided by exploiting the better 
signal-over-background level of the kinematic regions where the 
weak bosons have high transverse momenta~\cite{Butterworth:2008iy}. 
These channels are also important contributors to Higgs boson searches at
CMS~\cite{cms:h2bb2011} and the
Tevatron~\cite{tevatron:higgs_comb2011}.

This Letter presents searches in the \zhtollbb, \whtolnubb and
\zhtovvbb channels, where $\ell$ is either an electron or a muon. The
data used were recorded by the ATLAS experiment during the 2011 LHC
run at a centre-of-mass energy of $\sqrt{s}=7$~\tev\, and correspond
to integrated luminosities of $4.6$ to
$4.7$~\ifb~\cite{lumi11,lumi11conf}, depending on the analysis
channel. The leptonic decay modes of the weak bosons are selected to
suppress backgrounds containing only jets in the final state. In the
\zhtovvbb channel, the multijet background is suppressed by requiring
a large missing transverse energy.

  \section{The ATLAS Detector}

The ATLAS detector~\cite{atlas} consists of four main subsystems.  An
inner tracking detector is immersed in the 2~T magnetic field produced
by a superconducting solenoid. Charged particle position and momentum
measurements are made by silicon detectors in the
pseudorapidity\footnote{ATLAS uses a right-handed coordinate system
with its origin at the nominal interaction point (IP) in the centre of
the detector and the $z$-axis coinciding with the axis of the beam
pipe.  The $x$-axis points from the IP to the centre of the LHC ring,
and the $y$-axis points upward. Cylindrical coordinates ($r$,$\phi$)
are used in the transverse plane, $\phi$ being the azimuthal angle
around the beam pipe. The pseudorapidity is defined in terms of the
polar angle $\theta$ as $\eta = - \ln \tan(\theta/2)$.  For the
purpose of the fiducial selection, this is calculated relative to the
geometric centre of the detector; otherwise, it is relative to the
reconstructed primary vertex of each event.}  range $|\eta|<2.5$ and
by a straw tube tracker in the range $|\eta|<2.0$.  Calorimeters cover
$|\eta|<4.9$ with a variety of detector technologies. The liquid-argon
electromagnetic calorimeter is divided into barrel ($|\eta|<1.475$)
and endcap ($1.375<|\eta|<3.2$) sections.  The hadronic calorimeters
(using liquid argon or scintillating tiles as active materials)
surround the electromagnetic calorimeter and cover $|\eta|< 4.9$.  The
muon spectrometer measures the deflection of muon tracks in the field
of three large air-core toroidal magnets, each containing eight
superconducting coils. It is instrumented with separate trigger
chambers (covering $|\eta|<2.4$) and high-precision tracking chambers
(covering $|\eta|<2.7$).

  \section{Data and Monte Carlo Samples}
\label{sec:Samples}

The collision data used in this analysis are selected such that all
elements of the ATLAS detector were delivering high-quality data.  In
the \zhtollbb and the \whtolnubb analyses, events were primarily
collected using single-lepton triggers with a transverse momentum
(\pT) threshold of $20$~\gev\ for electrons, which was raised to
$22$~\gev\ as the instantaneous luminosity increased, and $18$~\gev\
for muons. In the \zhtollbb analysis, these triggers were supplemented
with a di-electron trigger with a threshold of $12$~\gev. The lepton
trigger efficiency is measured using a sample of $Z\rightarrow
\ell^+\ell^-$ events. The resulting efficiency, relative to the
offline selection, is close to $100\%$ for $ZH\rightarrow
e^+e^-b\bar{b}$ and $WH\rightarrow e\nu b\bar{b}$. It is around $95\%$
for the $ZH \rightarrow \mu^+\mu^-b\bar{b}$ channel and $90\%$ for the
$WH\rightarrow \mu\nu b\bar{b}$ channel, due to the lower angular
coverage of the muon trigger chambers with respect to the precision
tracking chambers. The missing transverse energy (\met) trigger used
for the \zhtovvbb channel has a threshold of $70\GeV$ and an
efficiency above $50\%$ for \met\ above $120\gev$. This efficiency
exceeds $99\%$ for \met\ above $170\gev$.  The efficiency curve is
measured in a sample of $W \rightarrow \mu\nu + \rm{jet}$ events
collected using muon triggers, which do not rely on the presence of
\met. The Monte Carlo (MC) simulation predicts the trigger efficiency
to be 5\% higher than that observed in collision data for $120\GeV \le
\MET < 160\GeV$ and agrees for $\MET \ge 160\GeV$. A correction
factor of $0.95\pm 0.01$ is therefore applied to the MC in the lower
$\MET$ region, and no trigger efficiency correction is applied
elsewhere.

The $WH$ and $ZH$ signal processes are modelled using MC events
produced by the {\sc Pythia}~\cite{pythia} event generator, interfaced
with the MRST modified leading-order (LO*)~\cite{Sherstnev:2007nd}
parton distribution functions (PDFs), using the AUET2B
tune~\cite{Aad:2011mctunes} for the parton shower, hadronization and
multiple parton interactions. The total cross sections for these
channels, as well as their corresponding uncertainties, are taken from
the LHC Higgs Cross Section Working Group
report~\cite{LHCHiggsCrossSectionWorkingGroup:2011ti}. Differential
next-to-leading order (NLO) electroweak corrections as a function of
the $W$ or $Z$ transverse momentum have also been
applied~\cite{Dittmaier:2012vm,Ferrera:2011bk}.  The Higgs boson decay
branching ratios are calculated with {\sc
Hdecay}~\cite{Djouadi:1997yw}.

The background processes are modelled with several different event
generators.  The {\sc
Powheg}~\cite{Alioli:2008tz,Nason:2009ai,Oleari:2011ey} generator, in
combination with MSTW 2008 NLO PDFs~\cite{Martin:2009iq} and
interfaced with the {\sc Pythia} program for the parton shower and
hadronization, is used to simulate $W + \ge 1 b$ jet events. The {\sc
Sherpa} generator~\cite{Gleisberg:2008ta} is used to simulate $Z$+$\ge
1 b$~jet and $Z$+$\ge 1 c$~jet events.  The {\sc Alpgen}
generator~\cite{alpgen} interfaced with the {\sc Herwig}
program~\cite{herwig} is used to simulate $W + \ge 1 c$ jet, $W + \ge
1$ light jet (i.e. not a $c$ or $b$ jet) and $Z + \ge 1$~light jet
events. The above background simulations include $\gamma^*$ production
and $Z/\gamma^*$ interference where appropriate.  The {\sc MC@NLO}
generator~\cite{mcatnlo}, using CT10 NLO PDFs~\cite{Lai:2010vv} and
interfaced to {\sc Herwig}, is used for the production of top-quarks
(single top and top-quark pair production).  The {\sc Herwig}
generator, is used to simulate the diboson ($ZZ$, $WZ$ and $WW$)
samples. The {\sc Herwig} generator uses the AUET2
tune~\cite{ATL-PHYS-PUB-2011-008} for the parton shower and
hadronization model, relies on MRST LO* PDFs (except for top
production) and is in all cases interfaced to {\sc Jimmy}~\cite{jimmy}
for the modelling of multiple parton interactions.  MC samples are
passed through the full ATLAS detector simulation~\cite{atlassim}
based on the {\sc Geant4}~\cite{GEANT4} program.

  \section{Reconstruction and Identification of Physics Objects}
\label{sec:ObjectSelection}

Events are required to have at least one reconstructed primary vertex
with three or more associated tracks with $\pT>0.4$~GeV in the inner
detector. If more than one vertex is reconstructed, the primary vertex
is chosen as the one with the highest sum of the squares of the
transverse momenta of all its associated tracks.

The charged leptons that are used to reconstruct the vector boson
candidate are required to satisfy $\pT>20\GeV$ in the \zhtollbb
channel, while this cut is increased to $\pT>25\GeV$ in the \whtolnubb
channel in order to be above the trigger threshold, and maintain a
high trigger efficiency. In both cases, the leptons must be central
($|\eta|<2.47$ for electrons and $|\eta|<2.5$ for muons) and have a
matching track in the inner detector that is consistent with
originating from the primary vertex.

Electron candidates are reconstructed from energy clusters in the
electromagnetic calorimeter and are required to pass identification
criteria based on the shower shape. Central electrons must have a
matching track in the inner detector that is consistent with
originating from the primary vertex and requirements are placed on
track quality and track-cluster
matching~\cite{atlas_egamma2010}. Further track and cluster related
identification criteria are applied to electron candidates in order to
reduce background from jets being misidentified as electrons.  The
criteria are tighter for $W$ decays, where the background is
larger. Muons are found by searching for tracks reconstructed in the
muon spectrometer with $|\eta|<2.7$. If the muon spectrometer track
matches a track in the inner detector, then the muon is reconstructed
from a combination of both tracks.

In order to suppress background from semileptonic heavy-flavour hadron
decays, the leptons are required to be isolated. In the \zhtollbb and
\whtolnubb channels the sum of the transverse momenta of all charged
tracks (other than those of the charged leptons) reconstructed in the
inner detector within a cone of $\Delta R =
\sqrt{(\Delta\eta)^2+(\Delta\phi)^2} <0.2$ from each charged lepton is
required to be less than 10\%~of the transverse momentum of the lepton
itself. In the \whtolnubb channel, the isolation requirement is
strengthened by requiring in addition that the sum of all transverse
energy deposits in the calorimeter within a cone of $\Delta R<0.3$
from the charged lepton be less than 14\% of the transverse energy of
the lepton itself.

 In order to suppress the top-quark background in the \zhtovvbb
channel, events containing electrons with $|\eta|<2.47$ and
$\pT>10\GeV$, or muons with $|\eta|<2.7$ and $\pT>10\GeV$ are removed.
Similar requirements are applied on any additional lepton
reconstructed in the \whtolnubb channel, but the minimum lepton $\pT$
is increased to $20\GeV$ if the additional lepton has the same charge
as, or a different flavour than the signal lepton.  Events with
forward electrons~\cite{atlas_egamma2010} ($2.47<|\eta|<4.5$) with
$\pT>20\GeV$ are also removed in the \whtolnubb channel.

Jets are reconstructed from energy clusters in the calorimeter using
the anti-$k_t$ algorithm~\cite{AntiKt} with a radius parameter of
$0.4$. Jet energies are calibrated using $\pt$- and $\eta$-dependent
correction factors based on MC simulation and validated with
data~\cite{JES}. A further correction is applied when calculating the
di-jet invariant mass, as described in section~\ref{sec:selection}
below. The contribution from jets originating from other collisions in
the same bunch crossing is reduced by requiring that at least $75\%$
of the summed transverse momentum of inner detector tracks (with $p_T >
0.4$~GeV) associated with the jet are compatible with originating from
the primary vertex.  Furthermore, a jet is required to have no
identified electron within $\Delta R \le 0.4$. Only jets with
$\pT>25$~\GeV\ and within the acceptance of the inner detector
($|\eta|<2.5$) are used to reconstruct Higgs boson candidates.  A
looser selection, for additional jets with $\pT>20$~\GeV\ and
$|\eta|<4.5$, is used to suppress additional hadronic activity in the
\whtolnubb channel.

Jets which originate from $b$ quarks can be distinguished from other
jets by the relatively long lifetime of hadrons containing $b$ quarks.
Such jets are primarily identified (``$b$-tagged'') by reconstructing
one or more secondary decay vertices from tracks within the jet, or by
combining the distances of closest approach to the primary event
vertex (impact parameters) of tracks in the
jet~\cite{btagnote2012c,btagnote2012l,btagnote2012b}. This information
is combined into a single discriminant $w$, such that a jet with
higher $w$ is more likely to be a $b$ jet. A selection cut on $w$ is
applied, resulting in an efficiency of about $70\%$ for identifying
true $b$ jets, with a $c$ jet rejection factor of about~$5$, and a
light jet rejection factor of about~$130$,
evaluated in simulated \ttbar\ events.

The missing transverse momentum and its magnitude are measured
from the vector sum of the transverse momentum vectors associated with
clusters of energy reconstructed in the calorimeters with
$|\eta|<4.9$~\cite{atlas_met3}. A correction is applied to the energy
of those clusters that are associated with a reconstructed physical
object (jet, electron, $\tau$-lepton, photon). Reconstructed muons are
also included in the sum, and any calorimeter energy deposits
associated with them are excluded.  To supplement the
calorimeter-based definition of \met\ in the \zhtovvbb\ channel,
the track-based missing
transverse momentum, \MPT, is calculated from the vector sum of the
transverse momenta of inner detector tracks associated with the
primary vertex~\cite{atlas_ptmiss}.

  \section{Event Selection}
\label{sec:selection}
 
Events in the \zhtollbb channel are required to contain exactly two
same-flavour leptons.  The two leptons must be oppositely charged in
the case of muons. This is not required for electrons since energy
losses from showering in material in the inner detector lead to a
higher charge misidentification probability.  The invariant mass of
the lepton pair must be in the range $83\gev<m_{\ell\ell}<99\gev$. A
requirement of $\met<50$~GeV reduces the background from top-quark
production.

Events in the \whtolnubb channel are required to contain a single
charged lepton and $\met>25$~GeV.  A
requirement on the transverse mass\footnote{The transverse
mass~($m_\mathrm{T}$) is defined from the transverse momenta and the
azimuthal angles of the charged lepton ($\pT^\ell$ and $\phi^\ell$)
and neutrino ($\pT^\nu$ and $\phi^\nu$): $m_{\mathrm{T}} =
\sqrt{2\pT^\ell\pT^\nu (1-\cos(\phi^\ell-\phi^\nu))}$, where $\pT^\nu
= \MET$.}  of $m_\mathrm{T}>40$~GeV is imposed to suppress the
multijet background.

The \zhtovvbb selection requires $\met>120$~GeV. A requirement of
\MPT$>30$~GeV is imposed to suppress events with poorly measured \met.
Cuts on the difference in azimuthal angle between the directions of
\met\ and $p_{\mathrm{T}}^{\mathrm{miss}}$, $\Delta \phi(\met ,
p_{\mathrm{T}}^{\mathrm{miss}}) < \pi/2$, and the difference in
azimuthal angle between \met\ and the nearest jet $\mathrm{min} (\Delta \phi (
\met, \mathrm{jet}) ) >1.8$ are applied to reduce the multijet
background, which is dominated by one or more jets being mismeasured
by the calorimeter.

The transverse momentum of the vector boson, $\pt^V$, is reconstructed
from the two leptons in the \zhtollbb channel, from the lepton and 
\met\ in the \whtolnubb channel and from \met\ in the \zhtovvbb channel.

Events in all channels are required to contain exactly two $b$-tagged
jets, of which one must have $\pT>45$~GeV and the other $\pT>25$~GeV.
If $\pt^V$ is less than $200$~GeV the two $b$-tagged jets are required
to have a separation of $\Delta R>0.7$, to reduce $W$+jet and
$Z$+jet backgrounds. Additionally, in the \zhtovvbb\ channel a cut on
the separation between the two jets of $\Delta R<2.0$ ($\Delta R<1.7$)
for $\pt^V<160$~GeV ($\pt^V>160$~GeV) is applied to reduce the
multijet background.  Events in the \zhtollbb channel may contain
additional non-$b$-tagged jets, while, in the \whtolnubb\ and
\zhtovvbb\ channels, events with additional jets are rejected, to
further suppress top-quark background. In the \whtolnubb analysis, where the
top-quark background is dominant, events containing additional jets with
$|\eta|<4.5$ and $\pT>20$~GeV are rejected, while in the \zhtovvbb\
channel the selection is restricted to jets with $|\eta|<2.5$ and
$\pT>25$~GeV.

In the \zhtovvbb\ analysis, further cuts are applied on the azimuthal
angle between $E_{\rm T}^{\rm miss}$ and 
the reconstructed transverse momentum of the $b\bar{b}$
system, $\Delta \phi(b\bar{b},\met)$,
to further reject multijet background. The cuts are $\Delta
\phi(b\bar{b},\met) > 2.7$ for $120 < \ptV < 160$~GeV and $\Delta
\phi(b\bar{b},\met) > 2.9$ for $\ptV \ge 160$~GeV.

A search for $H \to b\bar{b}$ decays is performed by looking for an
excess of events above the background expectation in the invariant
mass distribution of the $b$-jet pair~(\mbb). The value of the
reconstructed \mbb\ is scaled by a factor of 1.05, obtained from
MC-based studies, to account on average for e.g. losses due to soft
muons and neutrinos from $b$ and $c$ hadron decays.  To increase the
sensitivity of the search, this distribution is examined in bins of
$\pT^V$. As the expected signal is characterized by a relatively hard
$\pT^V$ spectrum, the signal to background ratio increases with
$\pT^V$.  The and \zhtollbb\ and \whtolnubb\ channels are examined in
four bins of the transverse momentum of the reconstructed $W$ or $Z$
boson, given by: $\ptV < 50$~GeV, $50 \le \ptV < 100$~GeV, $100 \le
\ptV <200$~GeV and $\ptV \ge 200$~GeV. In the \zhtovvbb\ search three
bins are defined: $120 < \ptV < 160$~GeV, $160 \le \ptV < 200$~GeV and
$\ptV \ge 200$~GeV.  The expected signal to background ratios for a
Higgs boson signal with $\mH=120$~GeV vary from about $1$\% in the
lowest $\ptV$\ bins to about $10$-$15$\% in the highest $\ptV$
bins. For this Higgs boson mass, 5.0\% and 2.4\% of the \zhtollbb\ and
\whtolnubb events are expected to pass the respective analysis
selections, with negligible contributions from other final states. On
the other hand, the \zhtovvbb\ analysis has a non-negligible
contribution from \whtolnubb~: 2.1\% of the \zhtovvbb\ signal and
0.2\% of the \whtolnubb\ signal are expected to pass the analysis
selection.

  \section{Background Estimation}
\label{sec:bkg}

Backgrounds are estimated using a combination of data-driven and
MC-based techniques.  Significant sources of background include top,
$W$+jet, $Z$+jet, diboson and multijet production.  The dominant
background in the \zhtollbb channel is $Z$+jet production. In the
\whtolnubb channel both the top-quark and $W$+jet
production are important.  In the \zhtovvbb channel, there is a
significant contribution from top, $W$+jet, $Z$+jet and diboson
production. Multijet production is a negligible background, except for
the \whtolnubb channel. 

The flavour composition of the $W$+jet and $Z$+jet backgrounds is
determined partially from data. 

The shapes of the \mbb\ distribution
of the top, $W$+jet and $Z$+jet backgrounds are taken from MC
simulation, with the respective normalizations being determined from
data. The ratio of single-top to top-pair production is taken from NLO
QCD computations~\cite{Aliev:2010zk}.  Multijet backgrounds are
estimated entirely from data. The diboson backgrounds are determined
from MC simulation with cross sections normalized to NLO QCD
computations~\cite{Campbell:1999ah,gg2WW}.

The flavour composition of the $W$+jet and $Z$+jet samples is
determined using templates produced from three exclusive MC samples
containing at least one true $b$ jet, at least one true $c$ jet, or
only light jets. The relative normalizations of the three components
are adjusted by fitting the distribution of the $b$-tagging
discriminating variable $w$ found in MC simulation to the distribution
found in control data samples dominated by $W$+jet and $Z$+jet
events. Once the relative normalizations of the flavour components
have been fixed, the overall normalizations are determined from data
in a separate step.

Sidebands in the \mbb\ distribution, defined by selecting events with
$\mbb<80\GeV$ or $150\GeV<\mbb<250\GeV$ along with the standard event
selection, are used to normalize the $Z$+jet, $W$+jet and top 
backgrounds.

In addition, two control regions which are 
dominated by top-quark production are used to further constrain 
the normalization of the top background.
The $ZH$ top control region selects events from the sidebands of the $Z$ boson mass peak:
$m_{\ell\ell} \in [60\GeV, 76\GeV] \cup [106\GeV, 150\GeV]$ with
$\met>50$~\gev, while the $WH$ top control region selects $W+3$ jet
events with two $b$-tagged jets. 

The normalizations of the $Z$+jet, $W$+jet and top-quark backgrounds
are determined in the \zhtollbb\ or \whtolnubb\ channels, by
simultaneous fits to the sidebands of the \mbb\ distributions, and
either the $ZH$ or $WH$ top control regions defined above.  In the
$WH$ sideband fit, the normalizations of the top-quark, the $W$+2 jet
and the $W$+3 jet distributions are varied. In the $ZH$ sideband fit,
the normalizations of the top-quark and $Z$+jet backgrounds are left
floating. The normalizations of the remaining sub-leading backgrounds
are left fixed in the fit at their expectation values from Monte Carlo
predictions, except for multi-jet production which is estimated from
data.  The relative data to MC normalization factors for top-quark
background agree with unity to within 20\% in both the \zhtollbb\ or
\whtolnubb\ sideband fits. The normalization of the top-quark
background in the \zhtollbb\ signal region is based on the $ZH$
sideband and control region fit result, while the normalization of the
same background in the \zhtollbb\ and \zhtovvbb\ signal regions is
based on the $WH$ sideband and control region fit result.  Monte Carlo
predictions are used to extrapolate the $Z$+jet ($W$+jet)
normalizations determined in the \zhtollbb\ (\whtolnubb) sidebands to
the signal regions of all three channels.  The normalization factors
for $W$+jet and $Z$+jet range from 0.8 to 2.4 depending on jet flavour
and multiplicity.  The MC to data normalization factors are applied to
several additional control samples with selections to enhance the $Z$,
$W$ or top-quark contributions. After these corrections are applied,
good agreement is found with the data in both shape and normalization
within the statistical and systematic uncertainties.

The backgrounds from multijet events are estimated entirely from
collision data.  For the \zhtollbb channel, the multijet background
normalization is determined from the sidebands of the $m_{\ell \ell}$
distribution in events containing at least two jets, and is found to
contribute less than 1\% and is therefore neglected. Multijet \MET\
templates for the \whtolnubb channel are obtained by selecting events
with lepton candidates failing the charged lepton analysis selection,
but satisfying looser lepton selections. The normalization is
determined by fitting these templates to the \MET\ distribution.  A
30\% uncertainty is determined from a comparison between the
normalized templates and the data in a multijet-dominated control
region, defined by requiring $\met\ <25$~\GeV\ and $m_\mathrm{T} <
40$~\GeV.

In the \zhtovvbb channel, the multijet background is estimated using
three control regions defined using two variables, $\Delta \phi(\met ,
p_{\mathrm{T}}^{\mathrm{miss}})$ and $\mathrm{min} ( \Delta \phi (
\met, \mathrm{jets}) )$, which showed no appreciable correlation. The
ratio of events with $\Delta \phi ( \met, \mathrm{jet} )>1.8$ to those
with $\mathrm{min} (\Delta \phi ( \met, \mathrm{jet} )) < 1.8$ is
determined for events with $\Delta \phi(\met ,
p_{\mathrm{T}}^{\mathrm{miss}})>\pi/2$. This ratio is then applied to
events with $\Delta \phi(\met , p_{\mathrm{T}}^{\mathrm{miss}})<\pi/2$
to estimate the multijet background in the signal region.  Upper
estimates of the multijet contamination in the signal region are found
to be 0.85, 0.04 and 0.26 events for $120 < \ptV < 160$~GeV, $160 \le
\ptV < 200$~GeV and $\ptV \ge 200$~GeV, respectively.  The accuracy of
the estimate is limited by the number of events in the control
regions.

The distribution of $m_{\ell\ell}$ in the \zhtollbb channel is shown
in Fig.~\ref{fig:control}(a) after all analysis requirements have been
applied (except for the di-lepton mass cut), including the requirement
of two $b$-tagged jets. The signal region is seen to be dominated by
$Z$+jet with smaller contributions from top-quark and diboson
production.  The \MET\ distribution in the \whtolnubb channel is shown
in Fig.~\ref{fig:control}(b) after all requirements, except for the
$m_{\mathrm{T}}$ and \met\ cuts. The signal region is seen to have
large contributions from top-quark production and $W$+jet, with
smaller contributions from the multijet background, $Z$+jet and
diboson production.  Figures~\ref{fig:control}(c) and
\ref{fig:control}(d) show the $\Delta \phi ( \met,
p_{\mathrm{T}}^{\mathrm{miss}})$ and $\mathrm{min}(\Delta \phi ( \met,
\mathrm{jet} ))$ distributions for the \zhtovvbb\ channel, after all
requirements except for those applied to these variables.  The
multijet background shape in Figure~\ref{fig:control}(c) is obtained
from data events with $\mathrm{min}(\Delta \phi ( \met, \mathrm{jet}
)) < 0.4$, after subtracting the remaining backgrounds, and normalized
to the data in the region defined by $\Delta \phi ( \met,
p_{\mathrm{T}}^{\mathrm{miss}}) > \pi/2$. In
Figure~\ref{fig:control}(d), the multijet shape is obtained from
events with $\Delta \phi ( \met, p_{\mathrm{T}}^{\mathrm{miss}}) >
\pi/2$ and normalized to data events with $\mathrm{min}(\Delta \phi (
\met, \mathrm{jet} )) < 0.4$.

 It can be
seen that the requirements on these variables effectively reduce the
multijet background.  The signal region has large contributions from
$Z$+jet and top, with smaller contributions from the $W$+jet, diboson
production and multijet backgrounds. For all distributions, the data
are reasonably well described by MC simulation and the multijet
background, which was determined from data.

\begin{figure*}[htb]
\begin{center}
\subfigure[]{\includegraphics[width=0.45\textwidth]{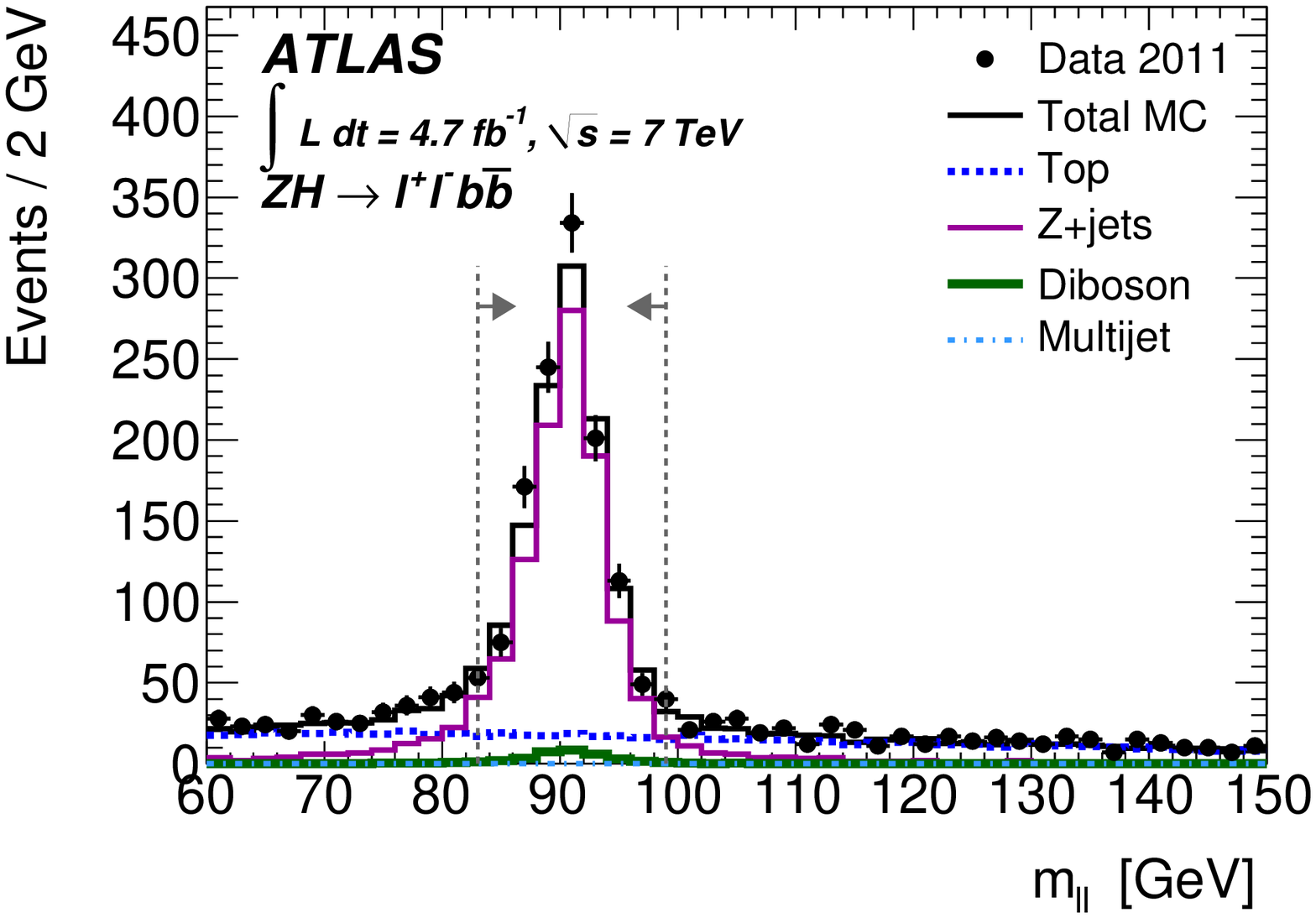}}
\subfigure[]{\includegraphics[width=0.45\textwidth]{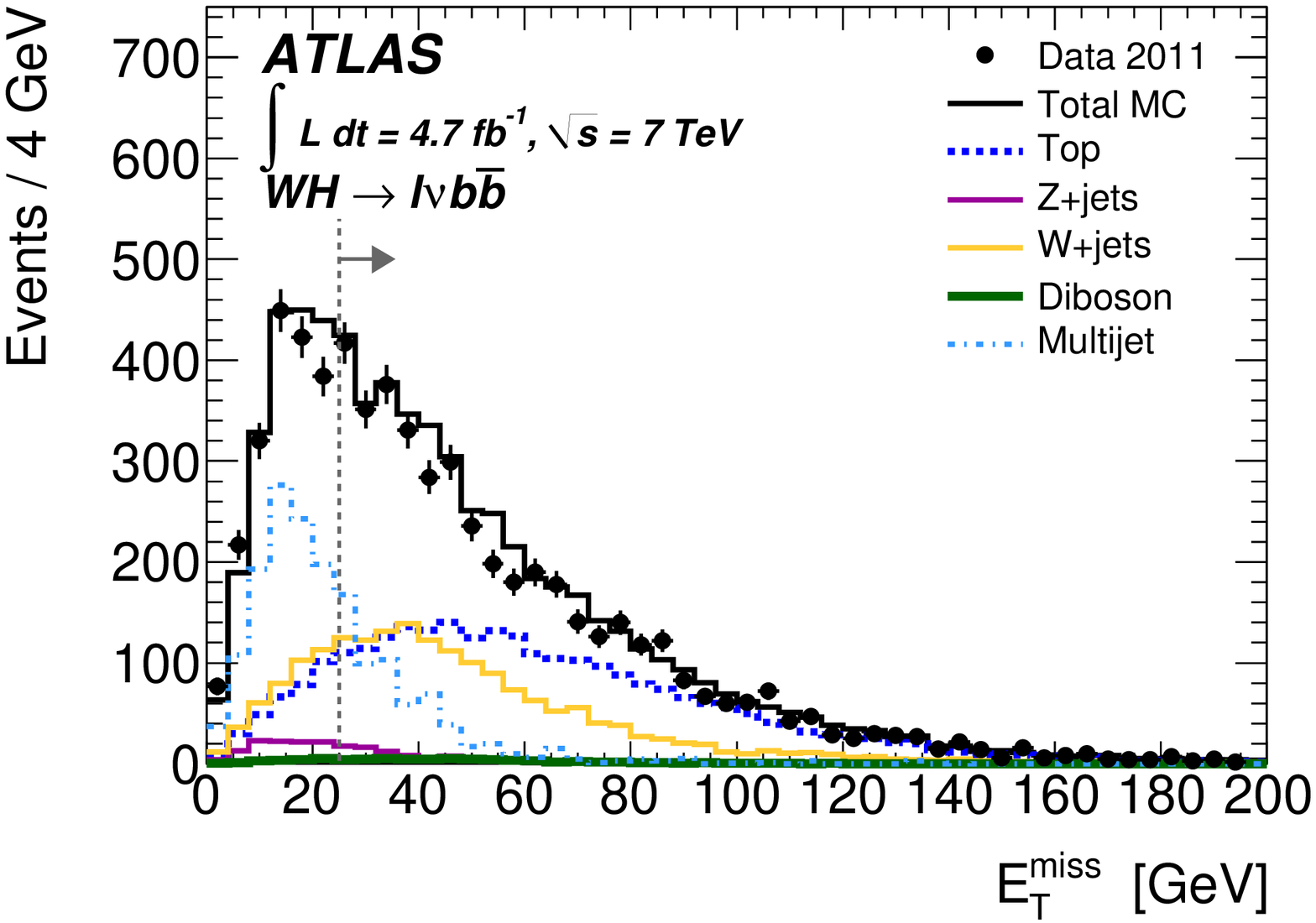}}
\subfigure[]{\includegraphics[width=0.45\textwidth]{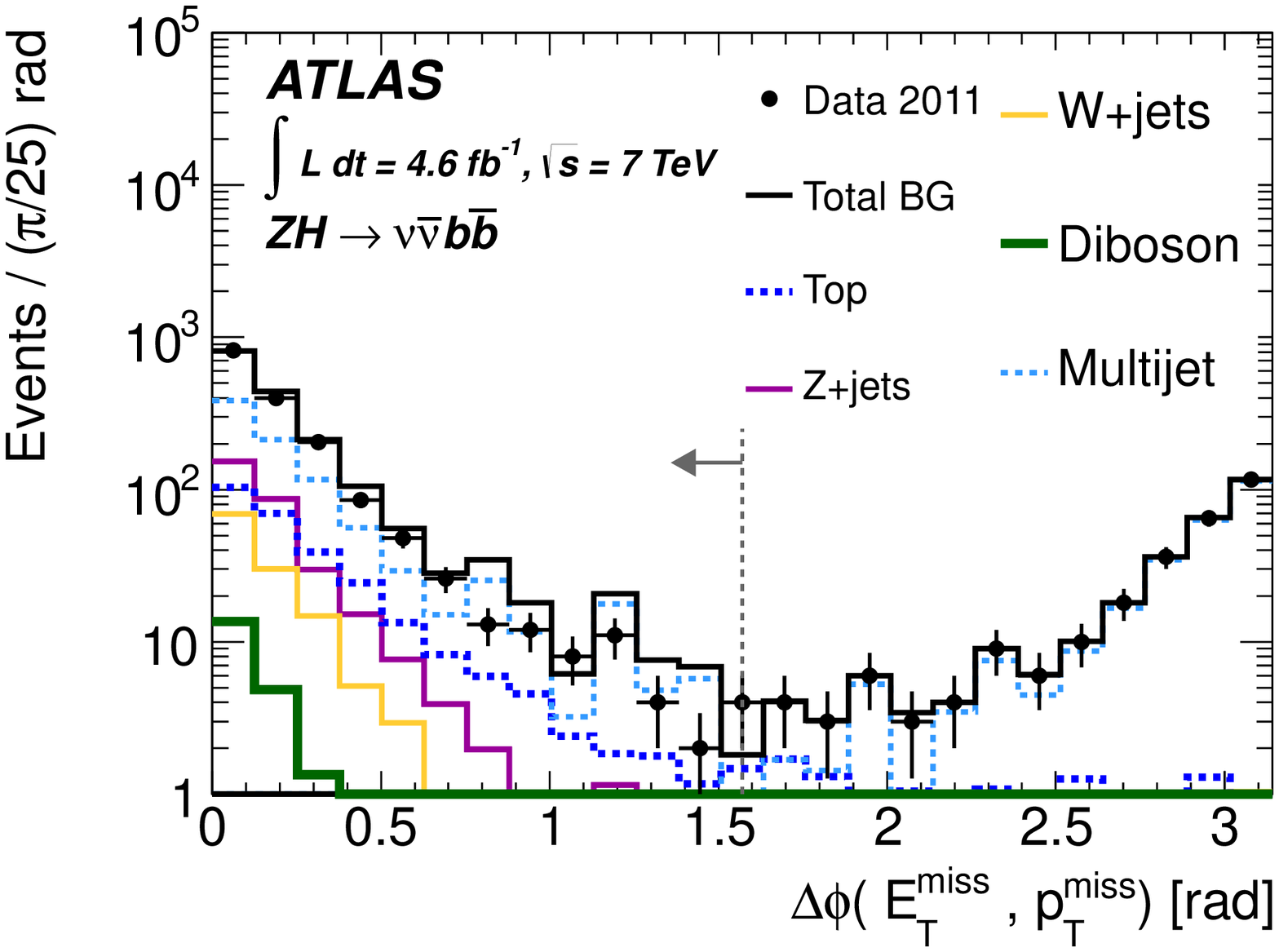}}
\subfigure[]{\includegraphics[width=0.45\textwidth]{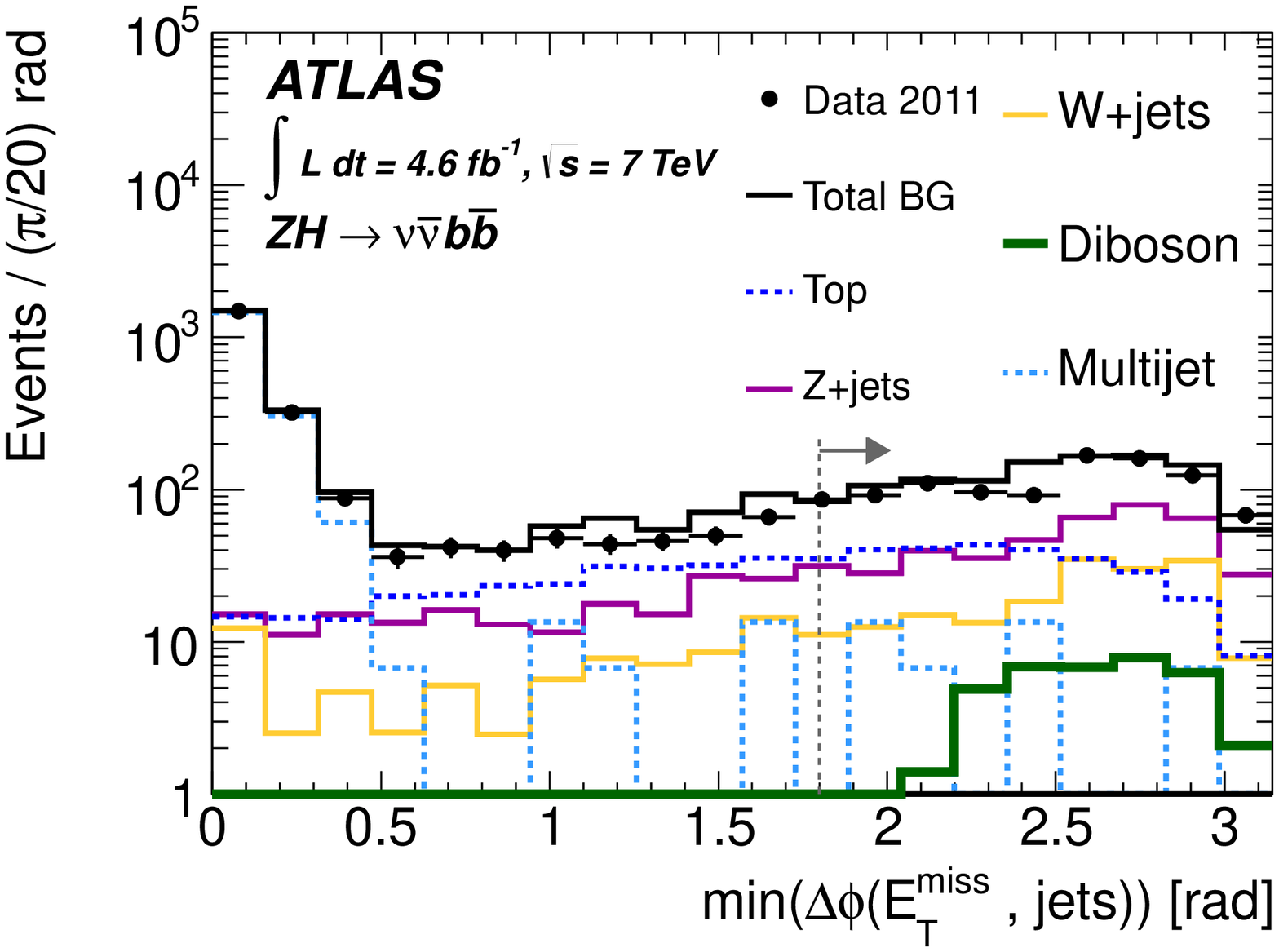}}
\caption{\label{fig:control} (a) The dilepton invariant mass
distribution in the \zhtollbb channel, (b) the missing transverse
energy without the $m_\mathrm{T}$ requirement in the \whtolnubb
channel, (c) the azimuthal angle separation between \met\ and
$p_{\mathrm{T}}^{\mathrm{miss}}$ and (d) the minimum azimuthal
separation between \met\ and any jet in the \zhtovvbb channel.  All
distributions are shown for events containing two $b$-tagged jets. The
various Monte Carlo background distributions are normalized to data
sidebands and control distributions and the multi-jet background is
entirely estimated from data as described in the text. The vertical
dashed lines correspond to the values of the cuts applied in each
analysis, and the horizontal arrows indicate the events selected by
each cut.}
\end{center}
\end{figure*}

  \section{Systematic Uncertainties}
\label{sec:systematics}

The sources of systematic uncertainty considered are those affecting
the various efficiencies (reconstruction, identification, selection),
as well as the momentum or energy of physics objects, the
normalization and shape of the \mbb\ distribution of the signal and
background processes, and the integrated luminosity.  Among these, the
leading instrumental uncertainties for all channels are related to the
uncertainty on the $b$-tagging efficiency, which varies between 5\%
and 19\% depending on the $b$-tagged jet $p_T$~\cite{btagnote2012b},
and the jet energy scale (JES) for $b$-tagged jets which varies
between 3\% and 14\% depending on the jet $p_T$ and
$\eta$~\cite{JESUpdate}. The $p_{T}$ dependence of the $b$-tagging
efficiency has been considered, based on the full covariance matrix of
the measured $b$-tagging efficiency in jet $p_{T}$
intervals~\cite{btagnote2012b}.  The uncertainty on the flavour
composition of the $Z+\jet$ and $W+\jet$ background is estimated by
varying the fraction of $Z+c$-jets and $W+c$-jets by 30\% as derived
from the fit described in Section~\ref{sec:bkg}.

 The uncertainties on the SM Higgs boson inclusive cross sections are
evaluated by varying the factorization and renormalization scales, and
by taking into account the uncertainties on the PDFs, on the strong
coupling constant and on the $H \to b\bar{b}$ branching fraction.
These uncertainties are estimated to be $\approx 4\%$ for both $WH$
and $ZH$ production and are treated according to the recommendations
given in
Refs.~\cite{LHCHiggsCrossSectionWorkingGroup:2011ti,Botje:2011sn,Ball:2011mu}.
Additional uncertainties are considered, as a function of the
transverse momentum of the $W$ and $Z$ bosons, which range from
$\approx 4\%$ to $\approx 8\%$, depending on channel and on the
$p_T^W$ or $p_T^Z$ interval. These correspond to the difference
between the inclusive and differential electroweak
corrections~\cite{Dittmaier:2012vm,Ferrera:2011bk}, and to differences
in acceptance between the {\sc Pythia} and {\sc Powheg}+{\sc Herwig}
generators. The latter arise mainly from the perturbative QCD model
uncertainty caused by rejecting events with three or more jets in the
\whtolnubb\ and \zhtovvbb\ analyses.

The uncertainties on the normalizations of the $Z$+jet, $W+$jet and
top-quark backgrounds are taken from the statistical uncertainties on
the fits to control regions and \mbb\ sidebands (see
Section~\ref{sec:bkg}) and from variations of the nominal fit result
induced by the remaining sources of systematic uncertainty.  The
resulting normalization uncertainties are applied to the \zhtovvbb\
channel.  A correlation between the normalizations of the $W$+jet and
top-quark backgrounds is introduced by the simultaneous fit to the
\mbb\ sidebands and the $WH$ top control region in the \whtolnubb\
channel. This correlation is taken into account when transferring to
the \zhtovvbb channel the uncertainties on the normalization of these
backgrounds.

The background normalization corrections are determined in an
inclusive way, using all selected events in the \zhtollbb and
\whtolnubb channels, and the shape of the \mbb\ and $\pt^V$
distributions are in each case taken from the MC
simulation. Therefore, a possible mismodelling of the underlying \mbb\
and $\pt^V$ distributions, as predicted by the MC generators, is also
considered. An uncertainty due to the shape of the \ptZ\ distribution
for the $Z$+jet background in the \zhtollbb channel is estimated by
finding variations of the MC \ptZ\ distribution in the \mbb\ sidebands
which cover any differences between MC simulation and data. The \mbb\
distribution of simulated $Z+$jet events is then reweighted according
to these variations, to estimate the effect on the final results. An
uncertainty due to the modelling of $W$+jet in the \whtolnubb channel
is estimated by reweighting the \ptW\ and \mbb\ distributions of
simulated $W+$jet events by variations motivated by a comparison of
different theoretical models ({\sc Powheg}+{\sc Pythia}, {\sc
Powheg}+{\sc Herwig}, {\sc aMC@NLO}+{\sc
Herwig}~\cite{Frederix:2011qg} and {\sc Alpgen}+{\sc Herwig}).
Theoretical uncertainties of $11\%$ and $15\%$ are applied to the
normalization of the diboson samples and the single-top sample,
respectively.  The normalization uncertainty for the multijet
background is taken to be $30\%$ for $WH \to \ell\nu b\bar{b}$, as
described in Section~\ref{sec:bkg}. For \zhtollbb and \zhtovvbb this
background is found to be negligible. The uncertainty in the
integrated luminosity has been estimated to be
$3.9\%$~\cite{lumi11,lumi11conf}.  This uncertainty is applied only to
backgrounds for which the normalization is not taken directly from a
comparison between data and MC simulation.  Where it is applied, this
systematic uncertainty is assumed to be correlated among the different
backgrounds.

  \section{Results}
\label{sec:Results}

The analysis is performed for five Higgs boson mass hypotheses between
$110$~\GeV\ and $130$~\GeV and the signal hypothesis is tested based
on a fit to the invariant mass distribution of the $b$-jet pair, \mbb,
in the signal region ($80 < \mbb < 150\GeV$). The \mbb\ distribution
is shown in Figs.~\ref{fig:zh_bins_mbb_120} --
\ref{fig:zhvv_bins_mbb_120} for each channel, separately for different
ranges of $\pt^V$.  The data distributions are overlaid with the
expectations from the MC simulation and data-driven
backgrounds. Within the experimental uncertainty, the data show no
excess over the background expectation. The signal shape is dominated
by the experimental resolution on the jet energy measurement. The
\mbb\ resolution for signal events is about 16~\GeV\ on average.

\begin{figure*}[htb]
  \begin{center}
    \subfigure[]{\includegraphics[width=0.45\textwidth]{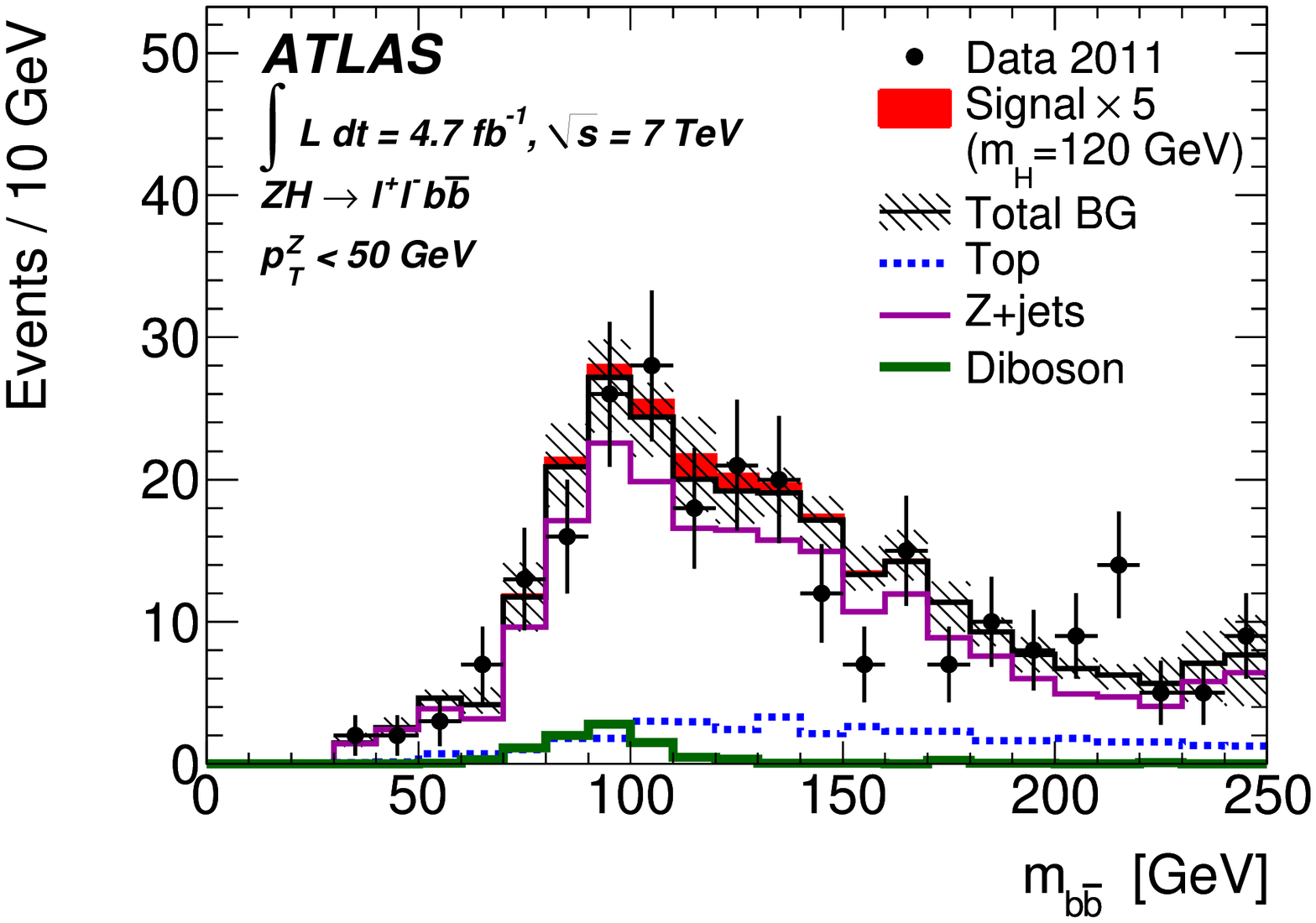}}
    \subfigure[]{\includegraphics[width=0.45\textwidth]{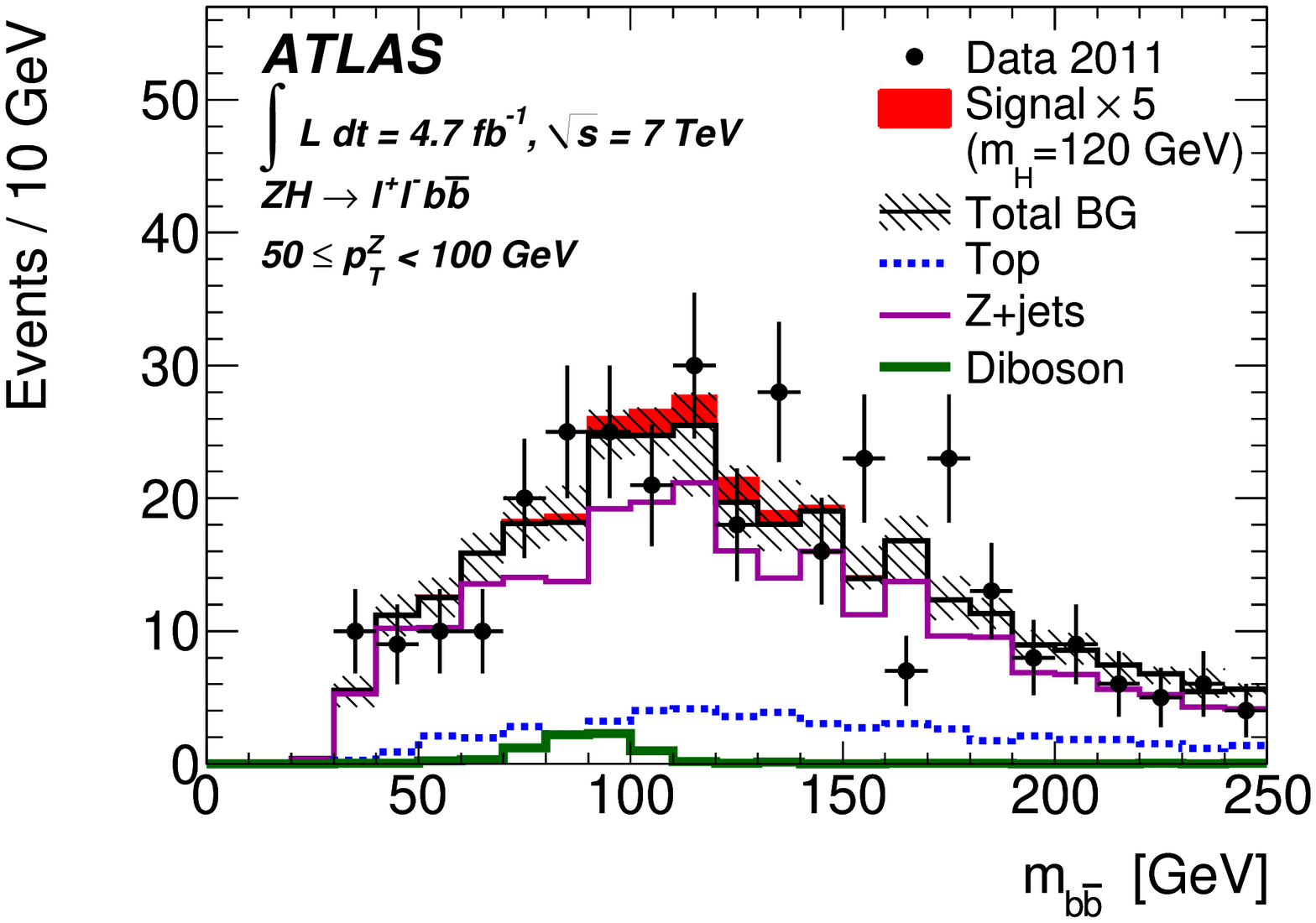}}
    \subfigure[]{\includegraphics[width=0.45\textwidth]{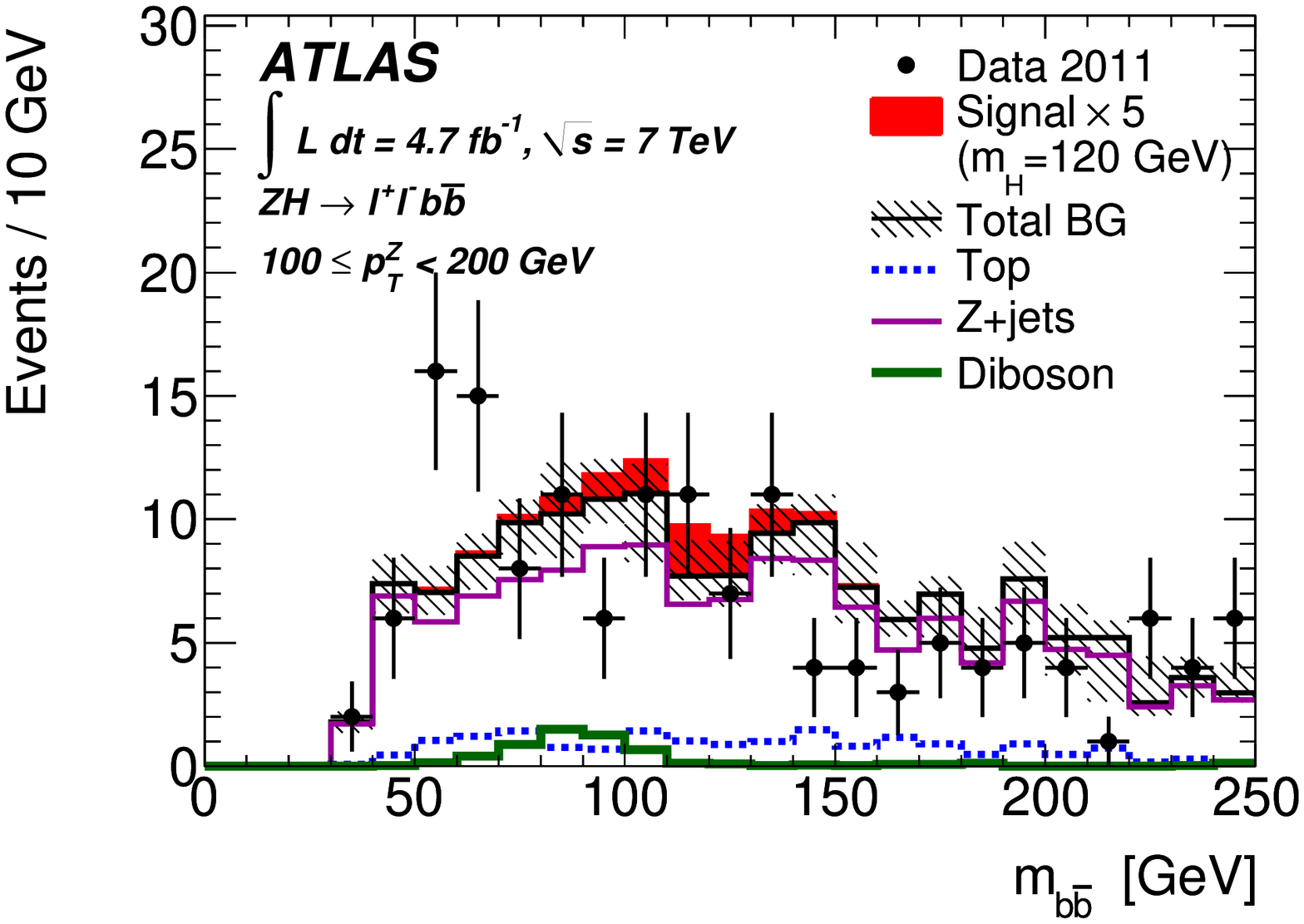}}
    \subfigure[]{\includegraphics[width=0.45\textwidth]{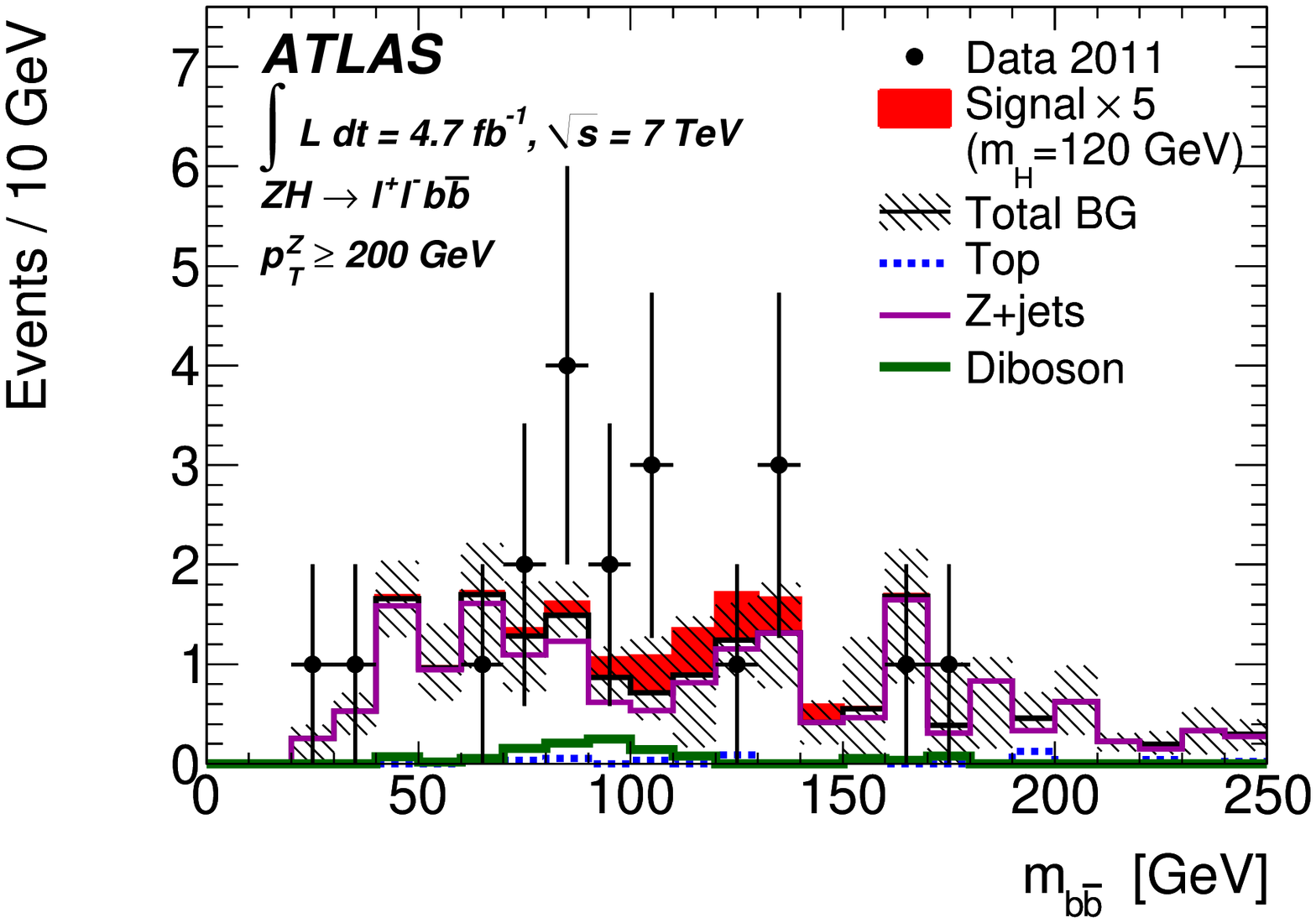}}
    \subfigure[]{\includegraphics[width=0.45\textwidth]{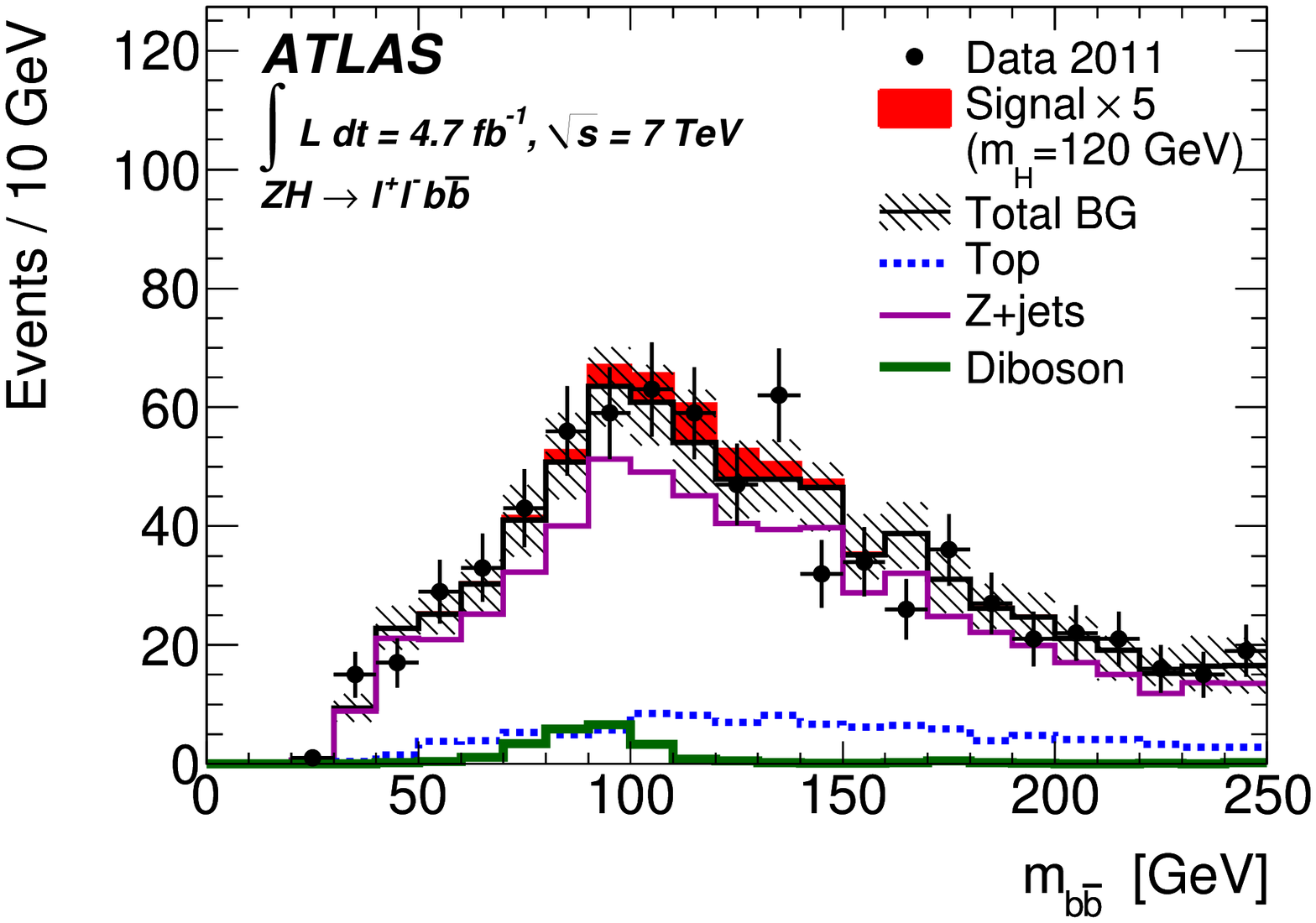}}
    \caption
	{ The invariant mass \mbb\ for \zhtollbb shown for the
	  different \ptZ\ bins: (a) $0 < \ptZ < 50\GeV$, (b) $50 \leq
	  \ptZ < 100\GeV$, (c) $100 \leq \ptZ < 200\GeV$, (d) $\ptZ
	  \geq 200\GeV$ and (e) for the combination of all \ptZ\ bins.
	  The signal distributions are shown for $\mH = 120\GeV$ and
	  are enhanced by a factor of five for visibility. The shaded
	  area indicates the total uncertainty on the background
	  prediction.  For better visibility, the signal histogram is
	  stacked onto the total background, unlike the various
	  background components which are simply overlaid in the
	  figure.  }
	\label{fig:zh_bins_mbb_120}
  \end{center}
\end{figure*}

\begin{figure*}[htb]
  \begin{center}
    \subfigure[]{\includegraphics[width=0.45\textwidth]{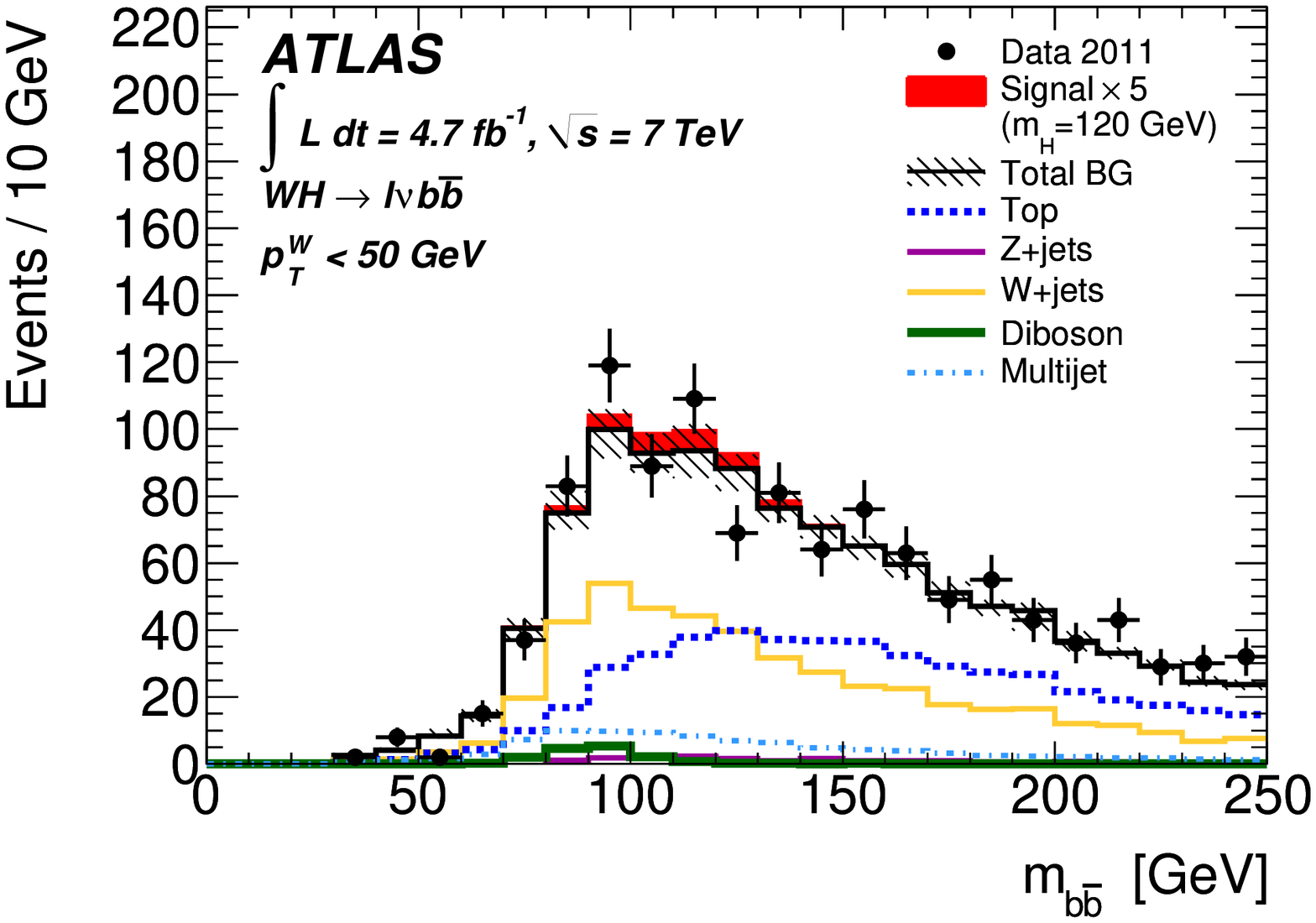}}
    \subfigure[]{\includegraphics[width=0.45\textwidth]{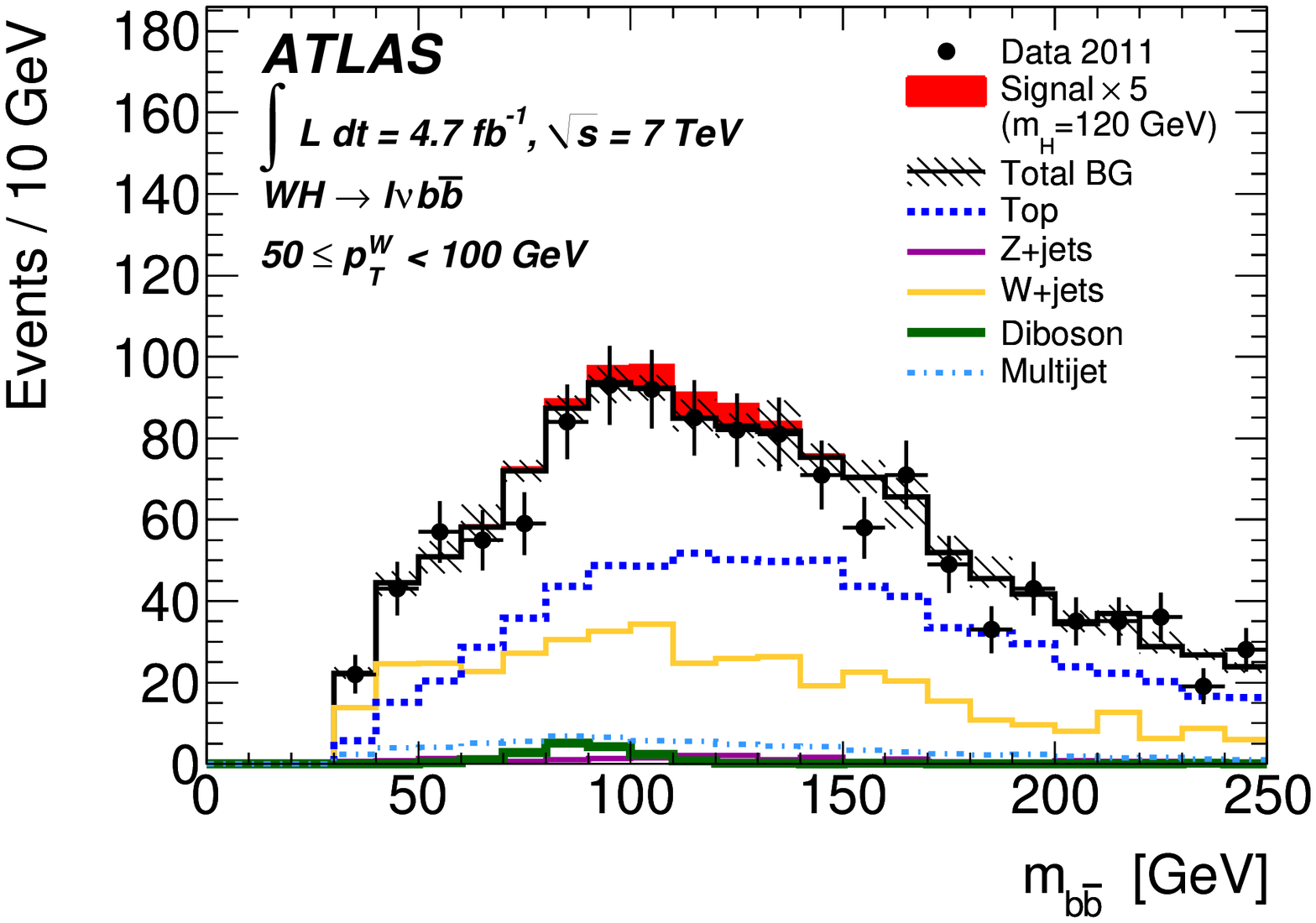}}
    \subfigure[]{\includegraphics[width=0.45\textwidth]{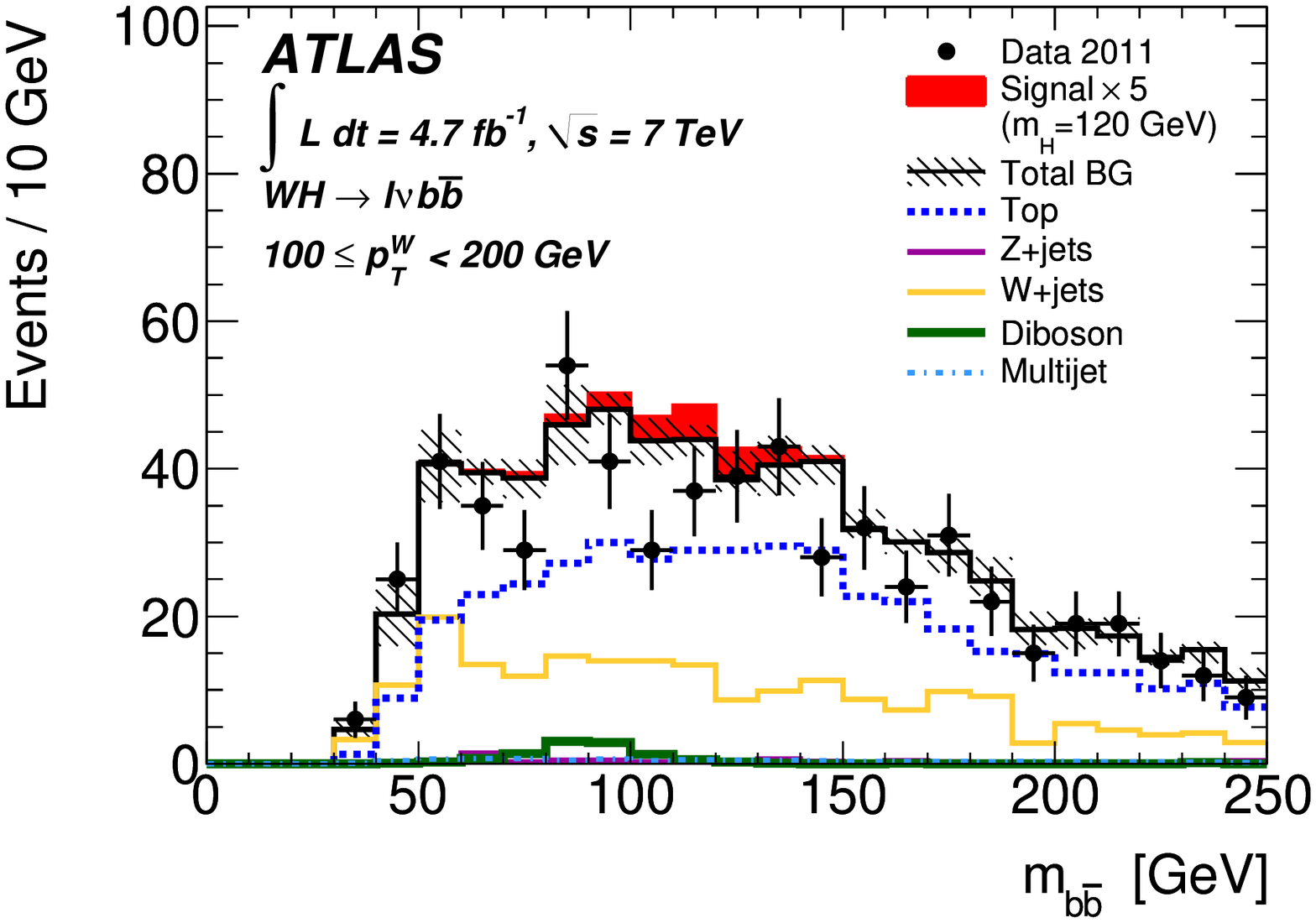}}
    \subfigure[]{\includegraphics[width=0.45\textwidth]{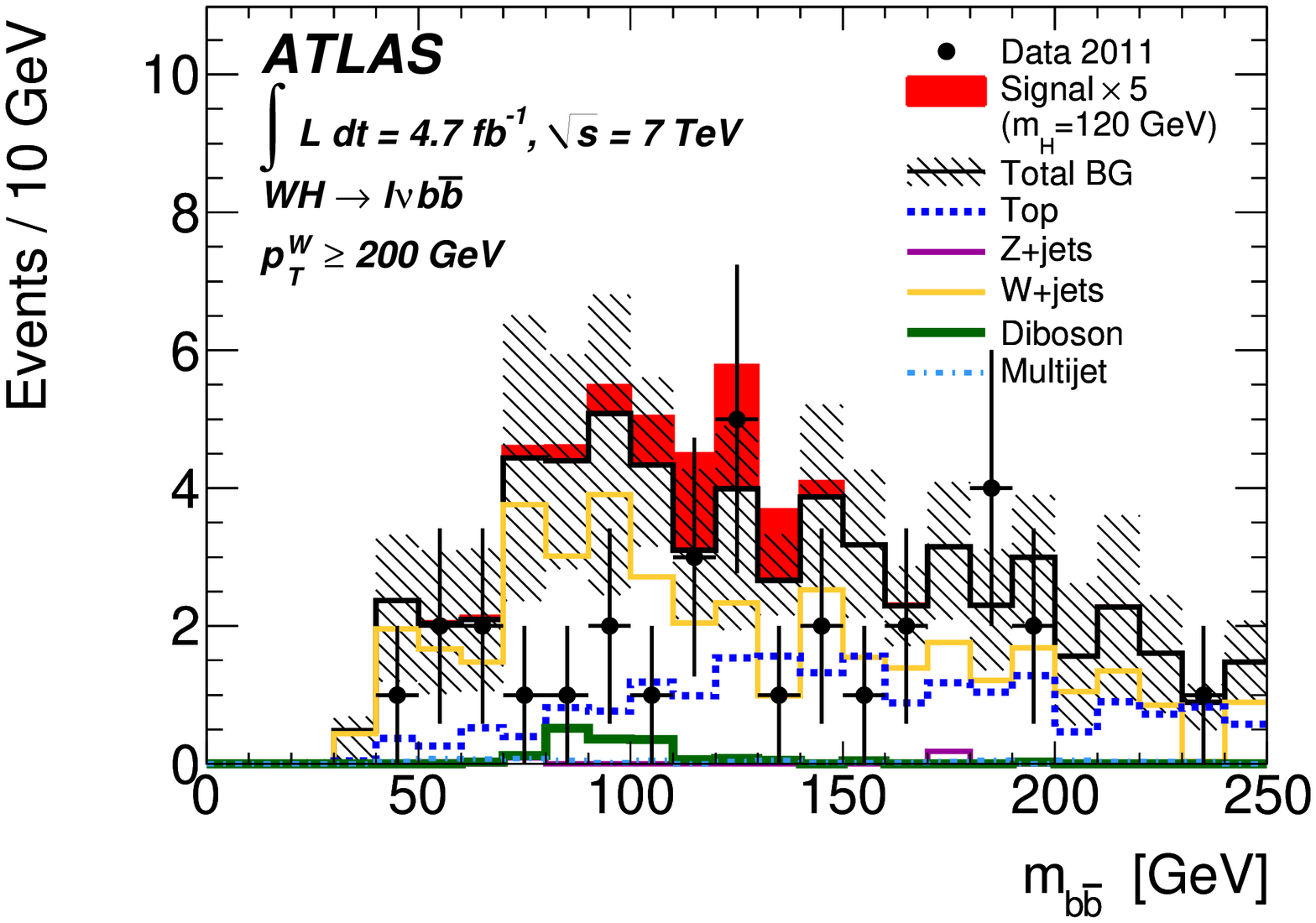}}
    \subfigure[]{\includegraphics[width=0.45\textwidth]{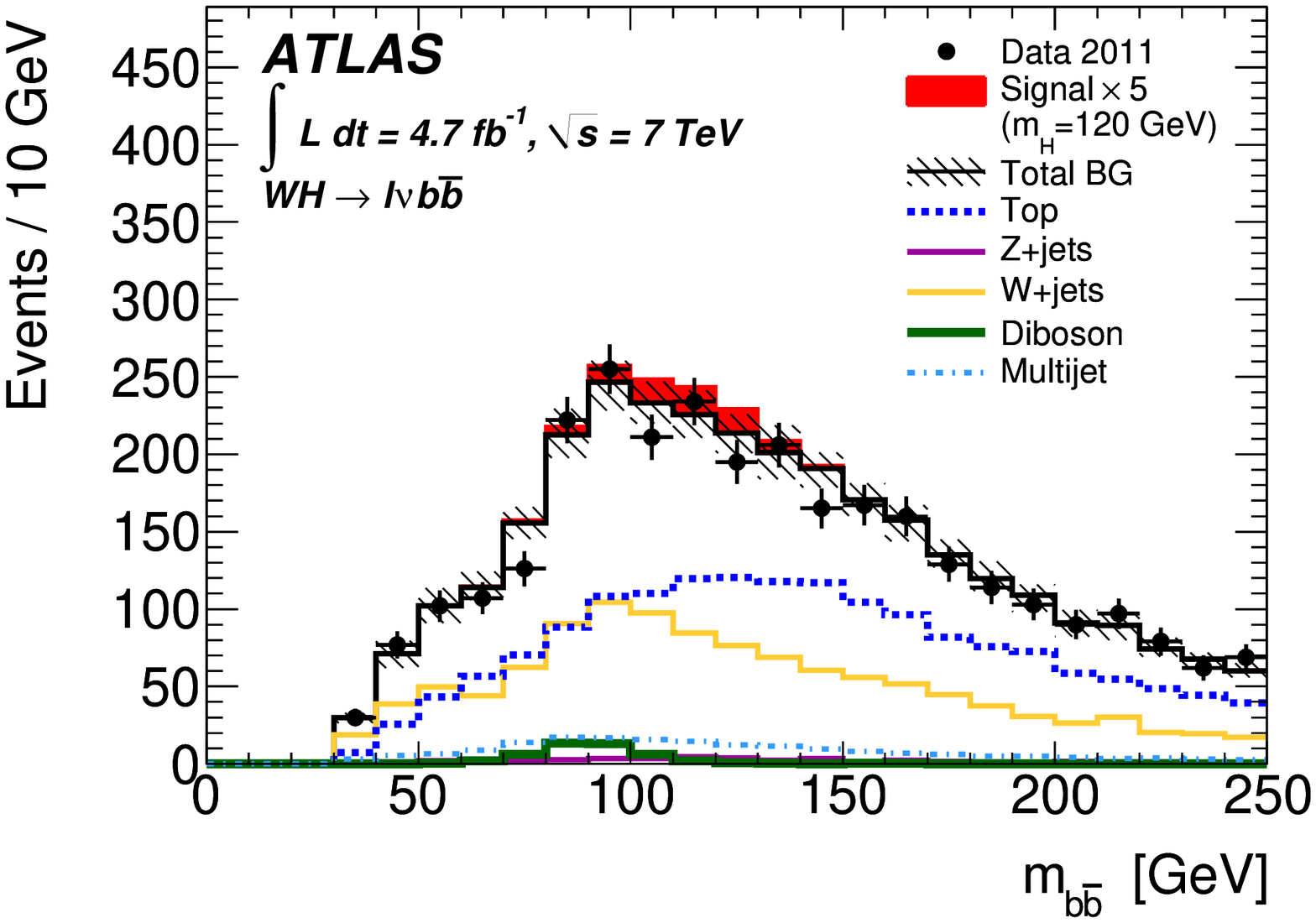}}
    \caption
	{ The invariant mass \mbb\ for \whtolnubb shown for the
	  different \ptW\ bins: (a) $0 < \ptW < 50\GeV$, (b) $50 \leq \ptW
	  < 100\GeV$, (c) $100 \leq \ptW < 200\GeV$, (d) $\ptW \geq
	  200\GeV$ and (e) for the combination of all \ptW bins.
	  The signal distributions are shown for $\mH = 120\GeV$ and
	  are enhanced by a factor of five for visibility. The shaded
	  area indicates the total uncertainty on the background
	  prediction.  For better visibility, the signal histogram is
	  stacked onto the total background, unlike the various
	  background components which are simply overlaid in the
	  figure. }
	\label{fig:wh_bins_mbb_120}
  \end{center}
\end{figure*}


\begin{figure*}[htb]
  \begin{center}
    \subfigure[]{\includegraphics[width=0.45\textwidth]{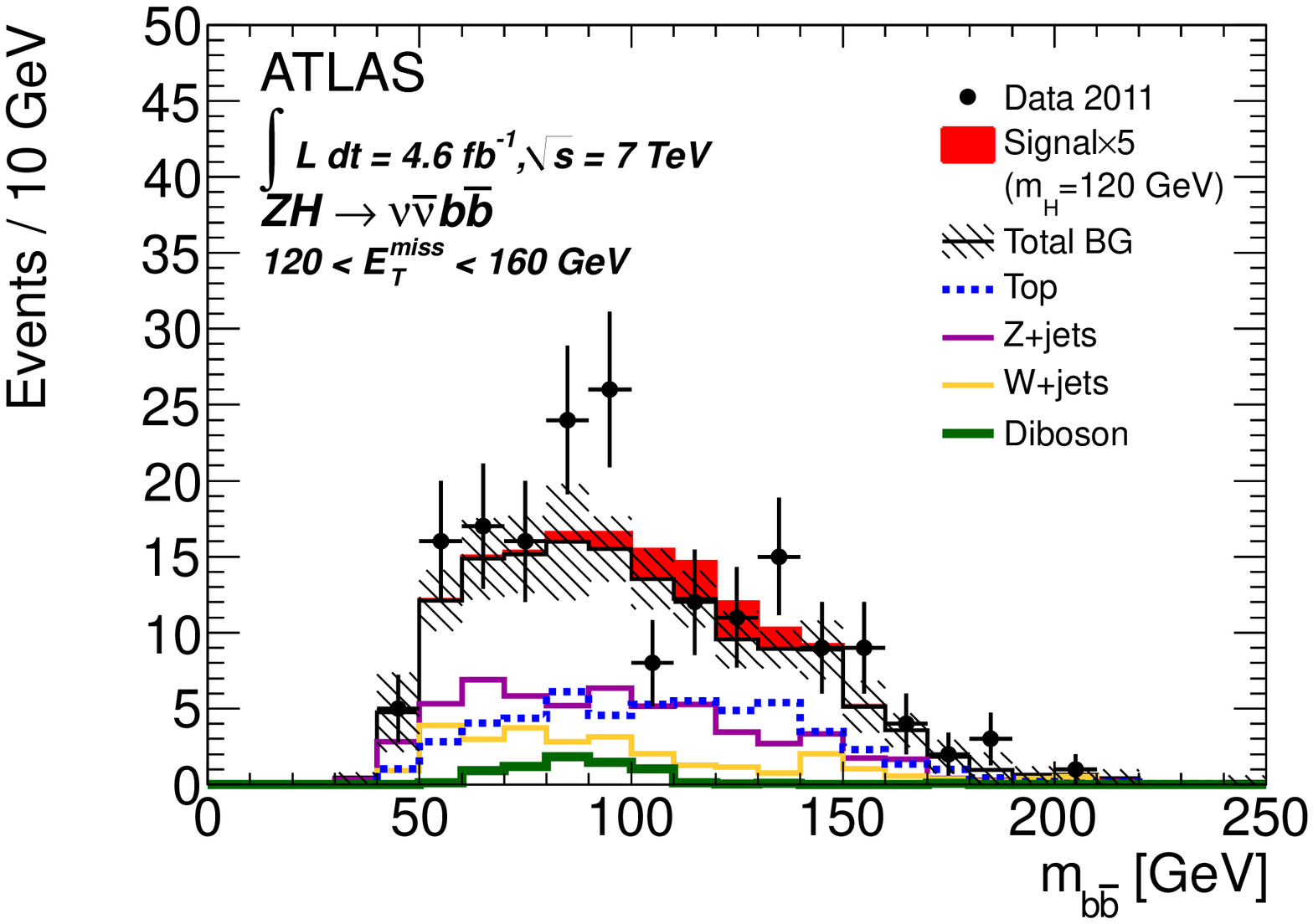}}
    \subfigure[]{\includegraphics[width=0.45\textwidth]{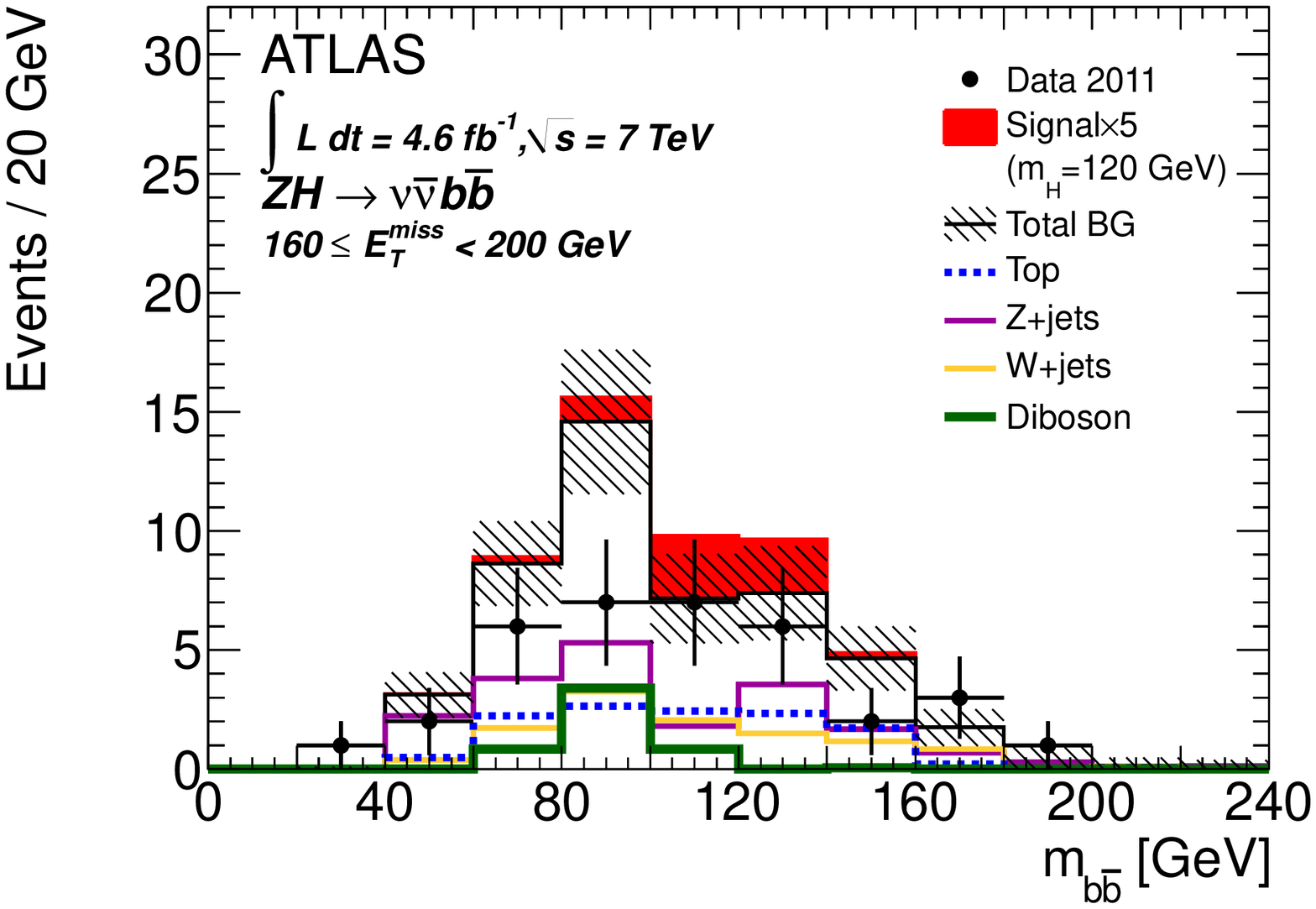}}
    \subfigure[]{\includegraphics[width=0.45\textwidth]{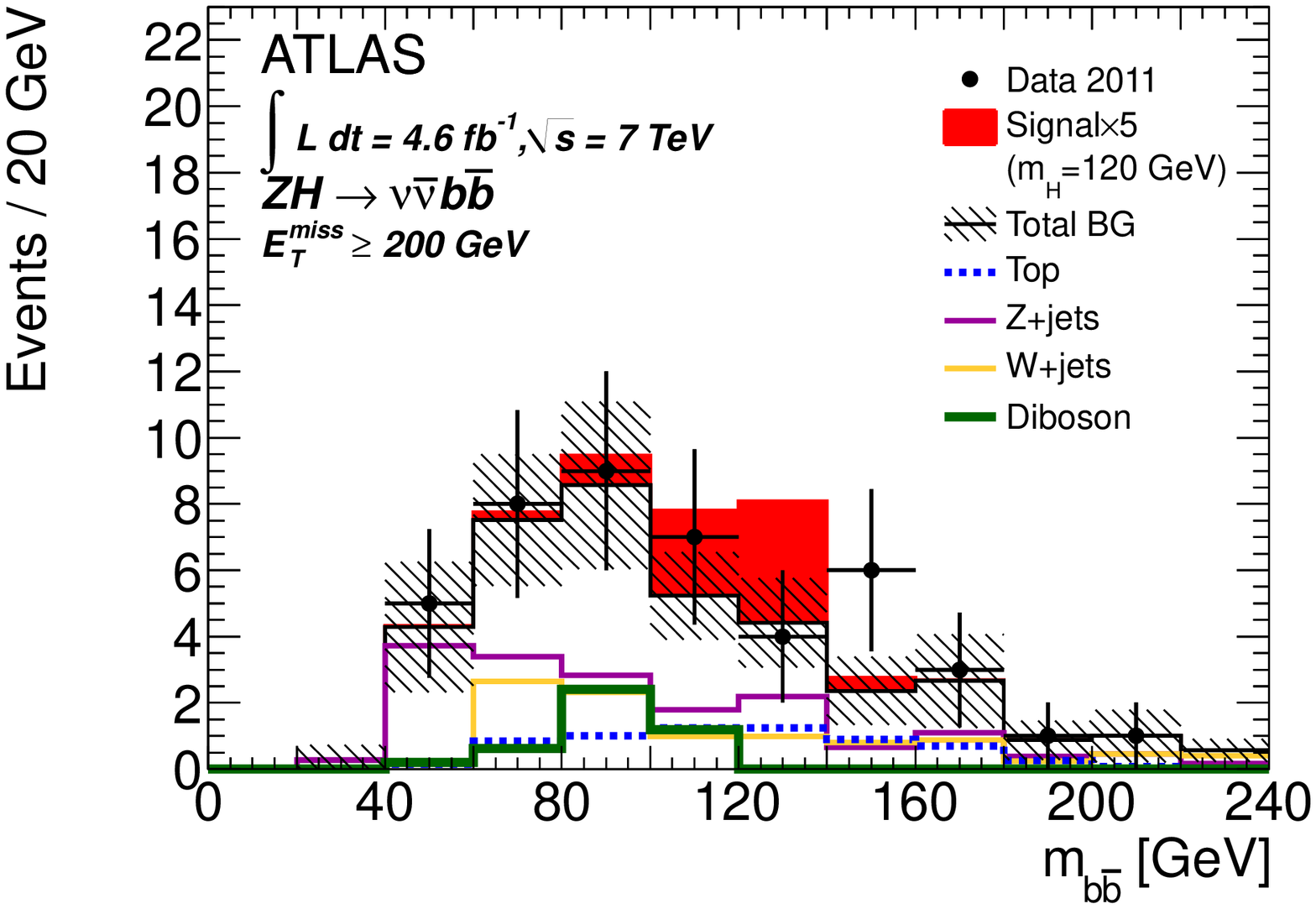}}
    \subfigure[]{\includegraphics[width=0.45\textwidth]{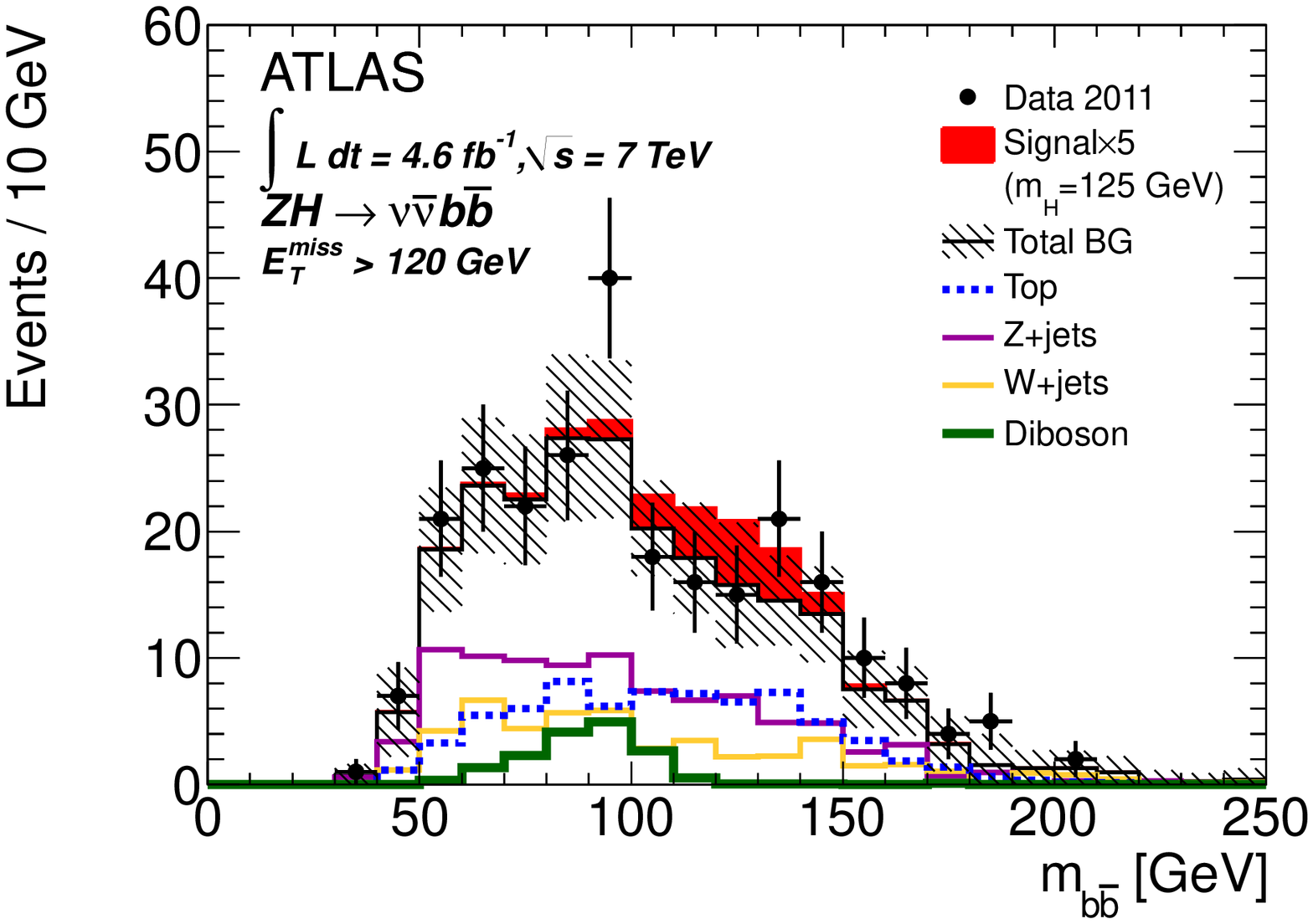}}

    \caption
	{ The invariant mass \mbb\ for \zhtovvbb shown for the
	  different \ptZ\ bins: (a) $120 < p_T^Z < 160\GeV$, (b) $160
	  \leq p_T^Z < 200\GeV$, (c) $p_T^Z \geq 200\GeV$ and (d) for
	  the combination of all $p_T^Z$ bins.  The signal
	  distributions are shown for $\mH = 120\GeV$ and are enhanced
	  by a factor of five for visibility. The shaded area
	  indicates the total uncertainty on the background
	  prediction. For better visibility, the signal histogram is
	  stacked onto the total background, unlike the various
	  background components which are simply overlaid in the
	  figure.
	  }
	\label{fig:zhvv_bins_mbb_120}
  \end{center}
\end{figure*}

The number of events in the signal region selected in data is
shown in Table~\ref{tab:nr_events} for each channel.  The expected
number of signal events for $\mH = 120\GeV$ is also shown, along with
the corresponding estimated number of background events.  Also shown
are the relative systematic uncertainties on the signal and total
background yields arising from the following sources: $b$-tagging
efficiency and mis-tag rate, background normalization, jet and \met\
uncertainties, lepton reconstruction and identification, integrated
luminosity, overlaid collision events (pileup), and uncertainties on
the MC predictions (theory). Uncertainties on
the shape of the \mbb\ distribution are also taken into account 
in the fit.

\begin{table*}
\centering
\caption{Number of data, simulated signal, and estimated background
    events in each bin of \ptV\ for the \whtolnubb, \zhtollbb and
    \zhtovvbb channels. The signal corresponds to a Higgs boson mass
    of $m_H=120\GeV$. The number of events is shown for the full
    signal region ($m_{b\bar{b}} \in [80\GeV, 150\GeV ]$). Background
    sources found to be negligible are signalled with ``--''. Relative
    systematic uncertainties on the hypothesized signal and estimated
    total background are shown.
    \label{tab:nr_events}}
	\small\addtolength{\tabcolsep}{-5pt}
	\begin{tabular}{|c | r| r| r| r| | r| r| r| r| | r| r| r|  }
	\hline
	& \multicolumn{4}{|c|}{\zhtollbb}& \multicolumn{4}{|c|}{\whtolnubb}& \multicolumn{3}{|c|}{\zhtovvbb}\\
	bin & \multicolumn{4}{|c|}{\ptV\ [GeV]}& \multicolumn{4}{|c|}{\ptV\ [GeV]}& \multicolumn{3}{|c|}{\ptV\ [GeV]}\\
	 & 0-50 & 50-100 & 100-200 & $>$200 & 0-50 & 50-100 & 100-200 & $>$200 & 120-160 & 160-200 & $>$200 \\
	\hline\hline
	\multicolumn{12}{|c|}{Number of events for $80 < \mbb < 150$ GeV}\\
	\hline\hline
	signal & 1.3 $\pm$ 0.1 & 1.8 $\pm$ 0.2 & 1.6 $\pm$ 0.2 & 0.4 $\pm$ 0.1 & 5.0 $\pm$ 0.6 & 5.1 $\pm$ 0.6 & 3.7 $\pm$ 0.4 & 1.2 $\pm$ 0.2 & 2.0 $\pm$ 0.2 & 1.2 $\pm$ 0.1 & 1.5 $\pm$ 0.2\\
	top & 17.4 & 24.1 & 7.3 & 0.2 & 229.9 & 342.7 & 201.3 & 8.2 & 35.2 & 8.3 & 4.1\\
	W+jets & -- & -- & -- & -- & 285.9 & 193.6 & 85.8 & 17.5 & 13.2 & 7.8 & 4.8\\
	Z+jets & 123.2 & 119.9 & 55.9 & 6.1 & 11.1 & 10.5 & 2.8 & 0.0 & 31.5 & 11.9 & 7.1\\
	diboson & 7.2 & 5.6 & 3.6 & 0.7 & 12.6 & 11.9 & 7.8 & 1.4 & 4.6 & 4.3 & 3.6\\
	multijet & -- & -- & -- & -- & 55.5 & 38.2 & 3.6 & 0.2 & -- & -- & --\\
	total BG & 148 $\pm$ 10 & 150 $\pm$ 6 & 67 $\pm$ 4 & 6.9 $\pm$ 1.2 & 596 $\pm$ 23 & 598 $\pm$ 16 & 302 $\pm$ 10 & 27 $\pm$ 5 & 85 $\pm$ 8 & 32 $\pm$ 3 & 20 $\pm$ 3 \\ \hline
	data & 141 & 163 & 61 & 13 & 614 & 588 & 271 & 15 & 105 & 22 & 25\\
	\hline
	\hline
	\multicolumn{12}{|c|}{Components of the relative systematic uncertainties of the background [\%]} \\
	\hline\hline
	$b$-tag eff  & 1.4 & 1.0 & 0.3 & 4.8 & 0.9 & 1.3 & 0.9 & 7.2 & 4.1 & 4.2 & 5.5\\
	BG norm  & 3.6 & 3.4 & 3.6 & 3.8 & 2.7 & 1.8 & 1.8 & 4.5 & 2.7 & 2.2 & 3.2\\
	jets/\met  & 2.1 & 1.2 & 2.7 & 5.1 & 1.5 & 1.4 & 2.1 & 9.5 & 7.7 & 8.2 & 12.1\\
	leptons  & 0.2 & 0.3 & 1.1 & 3.4 & 0.1 & 0.2 & 0.2 & 1.7 & 0.0 & 0.0 & 0.0\\
	luminosity  & 0.2 & 0.1 & 0.2 & 0.4 & 0.1 & 0.1 & 0.1 & 0.2 & 0.2 & 0.5 & 0.7\\
	pileup  & 0.9 & 1.6 & 0.5 & 1.3 & 0.1 & 0.2 & 0.8 & 0.5 & 1.6 & 2.5 & 3.0\\
	theory  & 5.2 & 1.3 & 4.7 & 14.9 & 2.2 & 0.3 & 1.6 & 14.8 & 2.9 & 4.0 & 7.7\\
	\hline
	total BG   & 6.9 & 4.3 & 6.6 & 17.3 & 3.9 & 2.7 & 3.4 & 19.6 & 9.7 & 10.6 & 16.0\\
	\hline\hline
	\multicolumn{12}{|c|}{Components of the relative systematic uncertainties of the signal [\%]} \\
	\hline\hline
	$b$-tag eff  & 6.4 & 6.4 & 7.0 & 13.7 & 6.4 & 6.4 & 7.0 & 12.1 & 7.1 & 8.2 & 9.2\\
	jets/\met  & 4.9 & 3.2 & 3.5 & 5.5 & 5.8 & 4.6 & 3.7 & 3.3 & 7.3 & 5.1 & 6.3\\
	leptons  & 0.9 & 1.2 & 1.7 & 2.6 & 3.0 & 3.0 & 3.0 & 3.2 & 0.0 & 0.0 & 0.0\\
	luminosity  & 3.9 & 3.9 & 3.9 & 3.9 & 3.9 & 3.9 & 3.9 & 3.9 & 3.9 & 3.9 & 3.9\\
	pileup  & 0.5 & 1.1 & 1.8 & 2.2 & 1.2 & 0.3 & 0.3 & 1.6 & 0.2 & 0.2 & 0.0\\
	theory  & 4.6 & 3.6 & 3.3 & 5.3 & 4.4 & 4.7 & 5.0 & 8.0 & 3.3 & 3.3 & 5.6\\
	\hline
	total signal   & 10.1 & 9.1 & 9.6 & 16.5 & 11.4 & 10.8 & 11.0 & 16.0 & 11.8 & 11.4 & 13.4\\
	\hline
	\end{tabular}
	\end{table*}

For each Higgs boson mass hypothesis, a one-sided upper limit is
placed on the ratio of the Higgs boson production cross section to its
SM value, $\mu = \sigma/\sigma_{\mathrm{SM}}$, at the 95\% confidence
level (CL).  The exclusion limits are derived from the
$CL_s$~\cite{cls_read} treatment of the $p$-values computed with the
profile likelihood ratio~\cite{asimov}, as implemented in the RooStats
program~\cite{roostats}, using the binned distribution of \mbb.  The
systematic uncertainties are treated by making the expected \mbb\
templates and sample normalizations dependent on additional fit
parameters (``nuisance parameters''), one for each systematic
uncertainty, which are then constrained with Gaussian terms within
their expected uncertainties. The dependence of the \mbb\ shapes on
the nuisance parameters is described with bin-by-bin linear
interpolation between the corresponding $+1\sigma$ or $-1\sigma$
variations and the nominal case.

The resulting exclusion limits are listed in Table~\ref{tab:limits}
for each channel and for the statistical combination of the three
channels. They are also plotted in Fig.~\ref{fig:exclusion}. The
limits are expressed as the multiple of the SM Higgs boson production
cross section which is excluded at 95\% CL for each value of the Higgs
boson mass.  The observed upper limits range between $7.7$ and $14.4$
for the \zhtollbb channel, between $3.3$ and $5.9$ for the \whtolnubb
channel and between $3.7$ and $10.3$ for the \zhtovvbb channel,
depending on the Higgs boson mass.  The combined exclusion limit for
the three channels together ranges from $2.5$ to $5.5$ times the SM
cross section, depending on the Higgs boson mass.  The limits include
systematic uncertainties, the largest of which arise from the top,
$Z$+jet, and $W$+jet background estimates, the $b$-tagging efficiency,
and the jet energy scale. The systematic uncertainties weaken the
limits by $25-40\%$ depending on the search channel.

\begin{figure*}
  \begin{center}
    \subfigure[]{\includegraphics[width=0.45\textwidth]{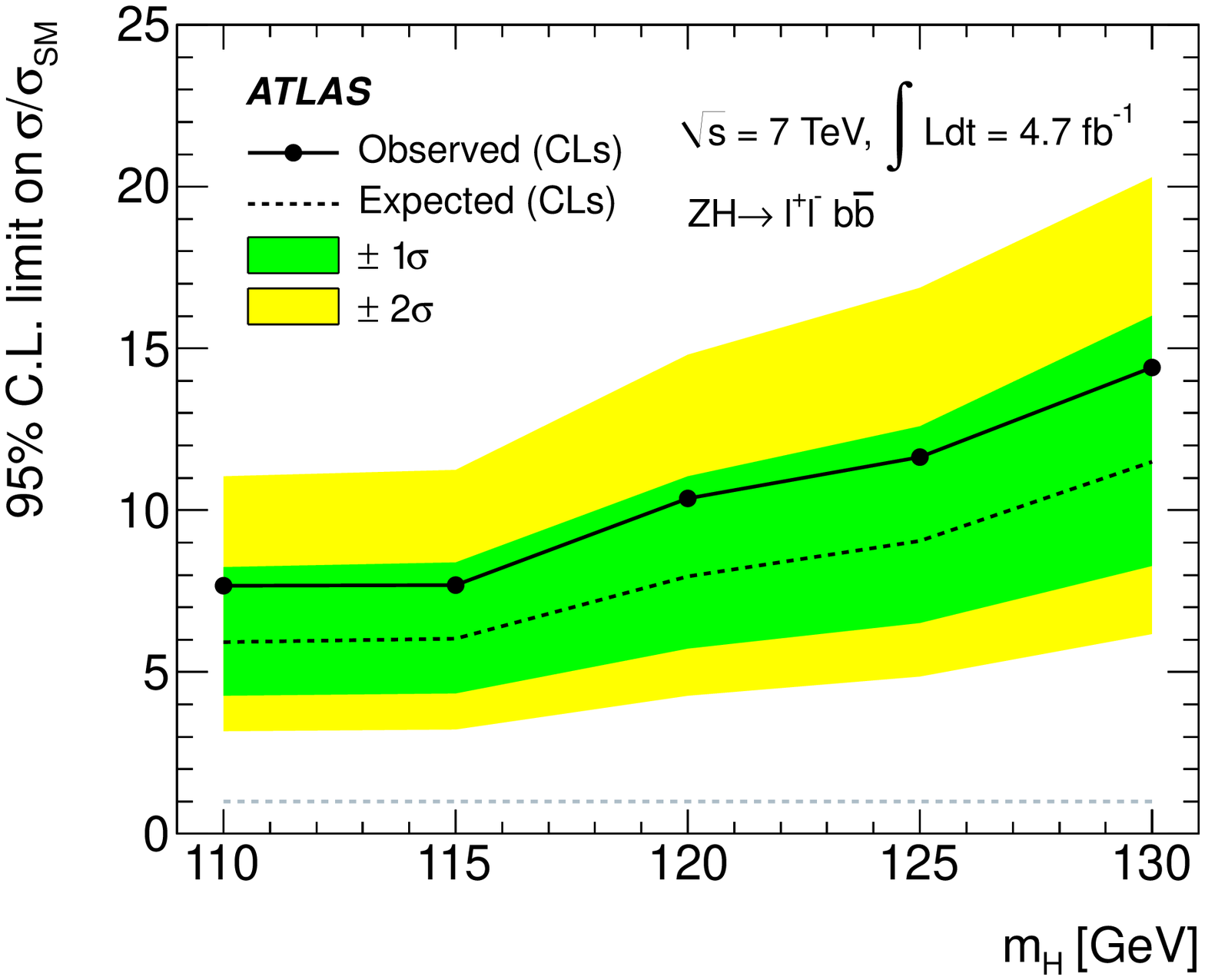}}
    \subfigure[]{\includegraphics[width=0.45\textwidth]{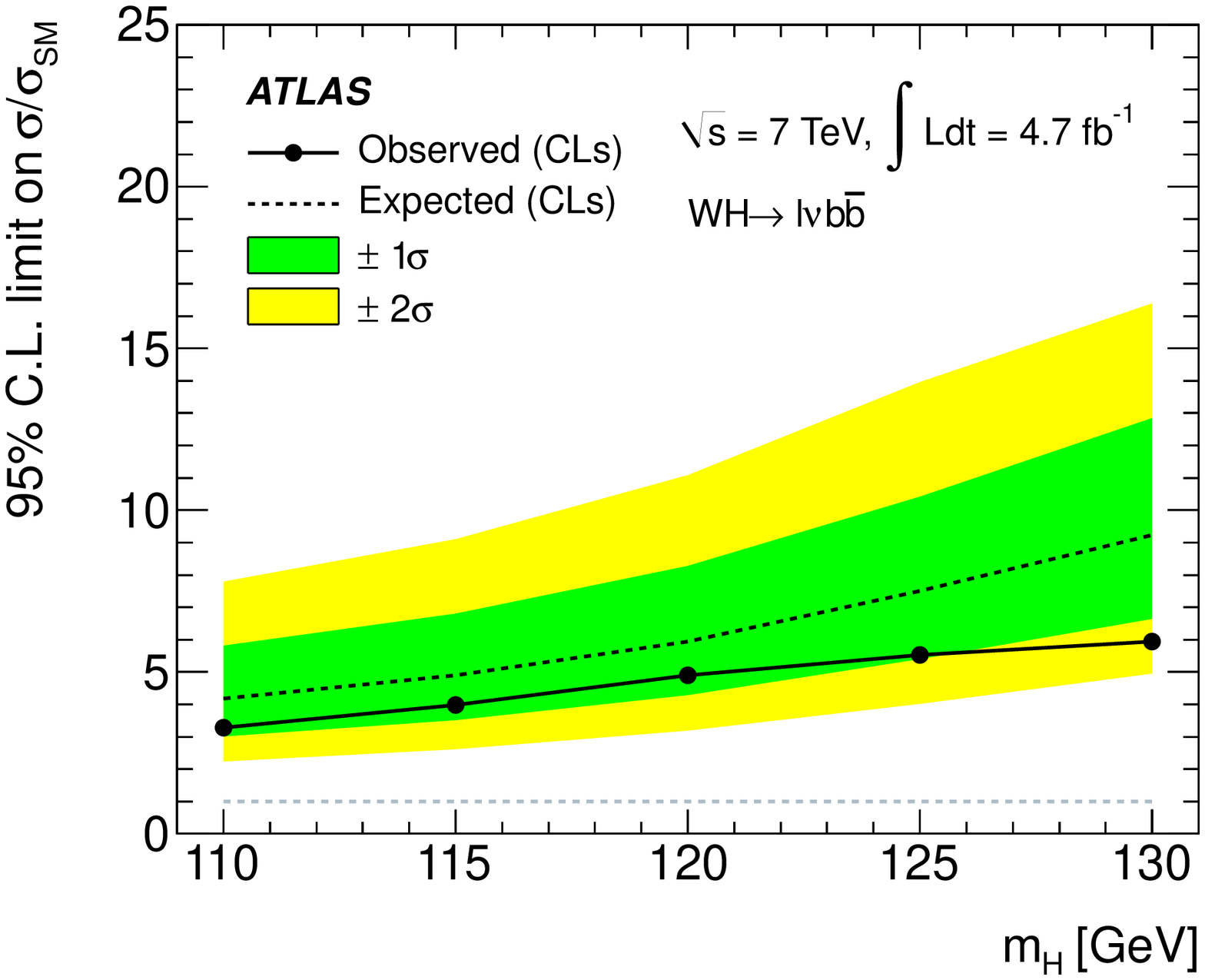}}
    \subfigure[]{\includegraphics[width=0.45\textwidth]{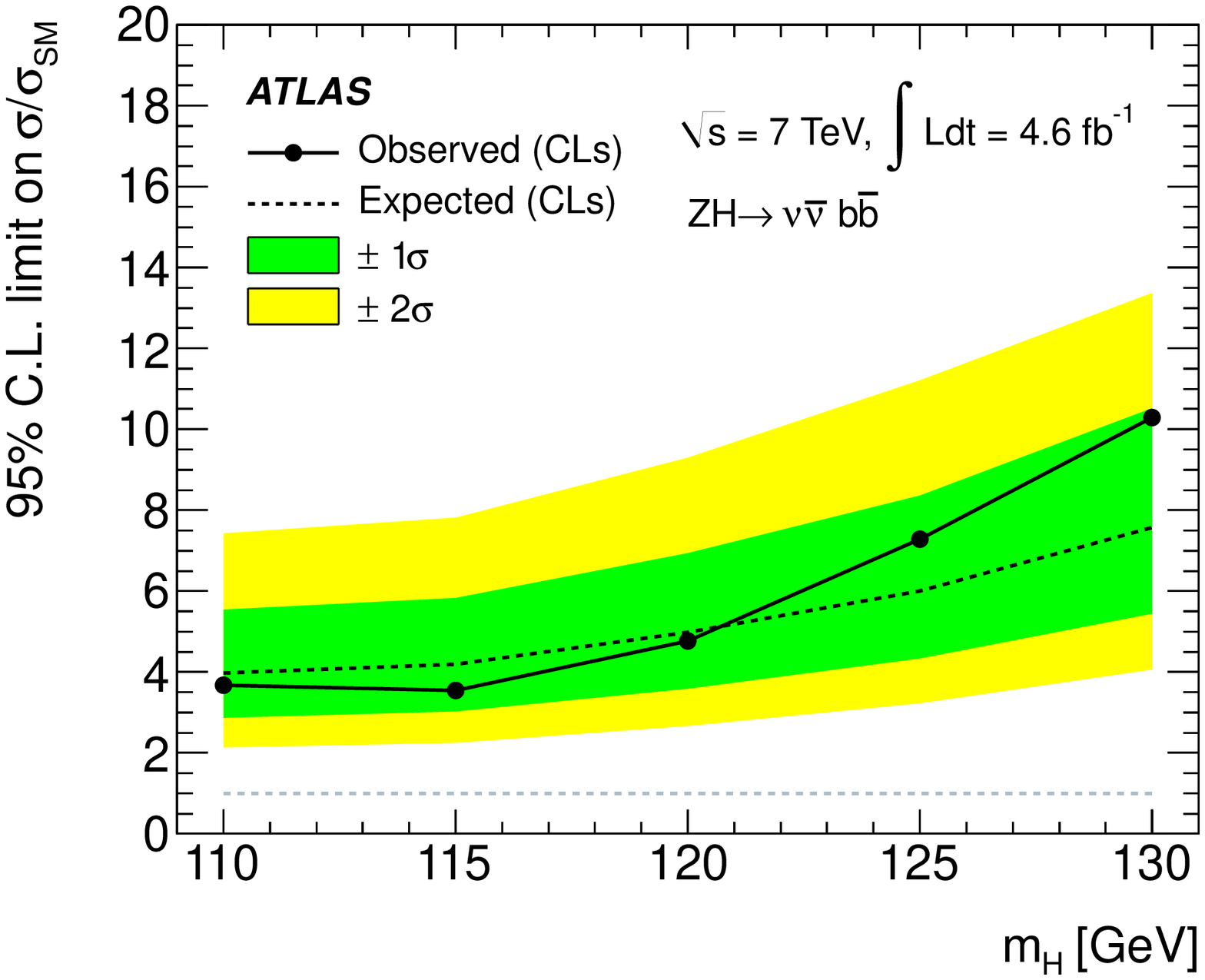}}
    \subfigure[]{\includegraphics[width=0.45\textwidth]{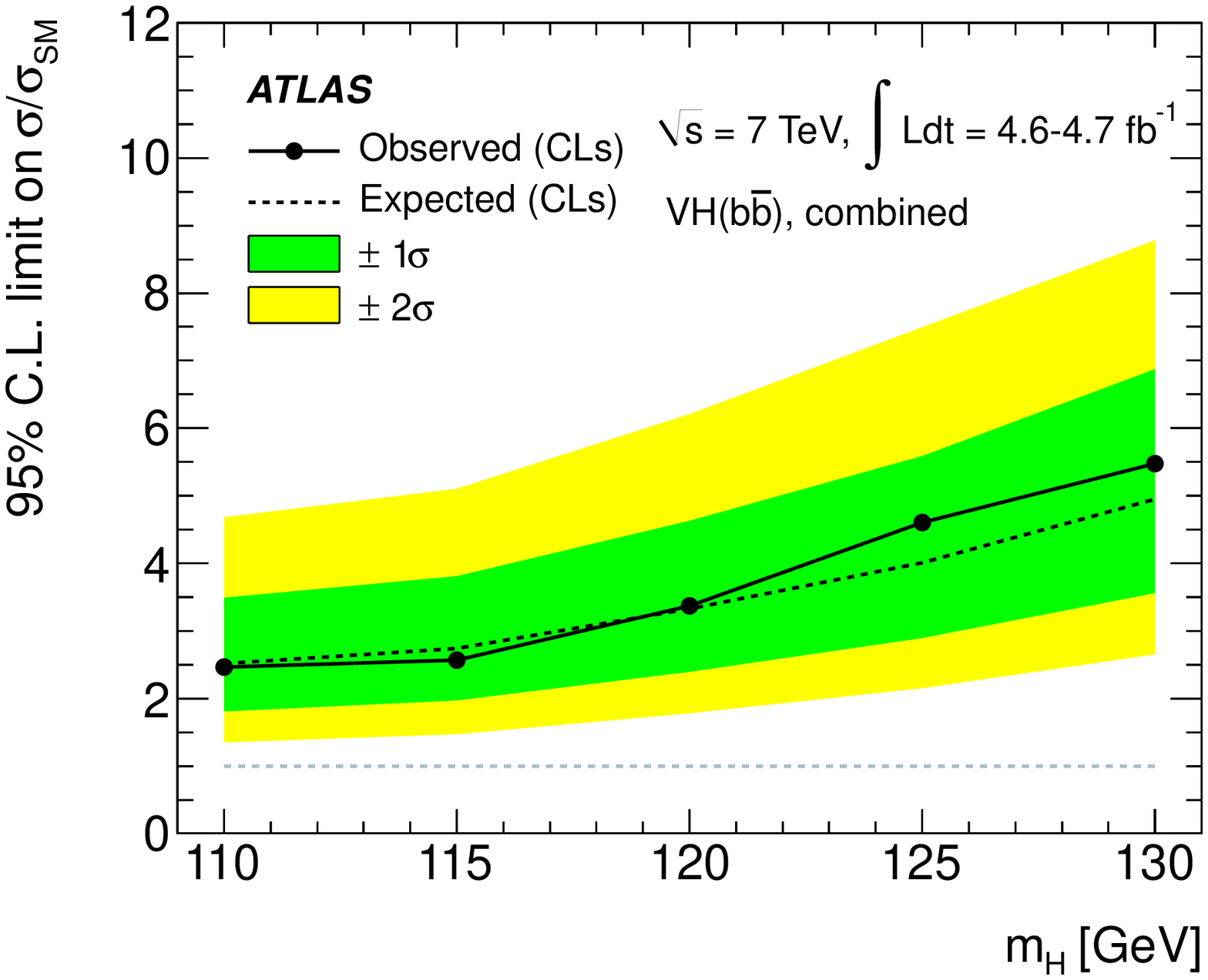}}
    \caption{Expected (dashed) and observed (solid line) exclusion
      limits for (a) the \zhtollbb, (b) \whtolnubb and (c) \zhtovvbb
      channels expressed as the ratio to the SM Higgs boson cross
      section, using the profile-likelihood method with $CL_s$.  The
      dark (green) and light (yellow) areas represent the 1$\sigma$
      and 2$\sigma$ ranges of the expectation in the absence of a
      signal. (d) shows the 95\% confidence level exclusion limits
      obtained from the combination of the three channels.
      \label{fig:exclusion}}
  \end{center}
\end{figure*}

\begin{table*}[htbp]
  \begin{center}
	  \caption{\label{tab:limits} The observed and expected $95\%$
            CL exclusion limits on the Higgs boson cross section for
            each channel, expressed in multiples of the SM cross
            section as a function of the hypothesized Higgs boson
            mass. The last two columns show the combined exclusion
            limits for the three channels.}
    \begin{tabular}{|c|c c|c c|c c|c c|}
      \hline 
      Mass  & \multicolumn{2}{|c|}{\zhtollbb} & \multicolumn{2}{|c|}{\whtolnubb} & \multicolumn{2}{|c|}{\zhtovvbb} & \multicolumn{2}{|c|}{Combined} \\
      $[ \mathrm{GeV} ]$  &         Obs. &         Exp.  &         Obs. &         Exp.  &         Obs. &         Exp.  &         Obs. &         Exp.   \\
      \hline      \hline 
      110 &          7.7 &          6.0 &            3.3 &          4.2 &            3.7 &          4.0 &          2.5 &          2.5  \\
      115 &          7.7 &          6.2 &            4.0 &          4.9 &            3.6 &          4.2 &          2.6 &          2.7  \\
      120 &         10.4 &          8.0 &            4.9 &          5.9 &            4.8 &          5.0 &          3.4 &          3.3  \\
      125 &         11.6 &          9.1 &            5.5 &          7.5 &            7.3 &          6.0 &          4.6 &          4.0  \\
      130 &         14.4 &         11.6 &            5.9 &          9.2 &           10.3 &          7.6 &          5.5 &          4.9  \\
      \hline 
    \end{tabular} 
  \end{center}
\end{table*}

  \section{Summary}
\label{sec:Summary}

This Letter presents the results of a direct search by ATLAS for the
Standard Model Higgs boson produced in association with a $W$ or $Z$
boson. The following decay channels are considered: $ZH \to
\ell^+\ell^- b\bar{b}$, $WH \to \ell\nu b\bar{b}$ and $ZH \to
\nu\bar{\nu} b\bar{b}$, where $\ell$ corresponds to an electron or a
muon.  The mass range $110 < \mH < 130\GeV$ is examined for five Higgs
boson mass hypotheses separated by 5~GeV steps.  The three channels
use datasets corresponding to $4.6$ -- $4.7$~\ifb\ of $pp$ collisions
at $\sqrt{s}=7$~TeV.  No significant excess of events above the
estimated backgrounds is observed. Upper limits on Higgs boson
production, at the $95\%$ confidence level, of $2.5$ to $5.5$ times
the Standard Model cross section are obtained in the mass range
$110-130\GeV$. The expected exclusion limits range between $2.5$ and
$4.9$ for the same mass interval.

\section{Acknowledgements}
  
We thank CERN for the very successful operation of the LHC, as well as
the support staff from our institutions without whom ATLAS could not
be operated efficiently.

We acknowledge the support of ANPCyT, Argentina; YerPhI, Armenia; ARC,
Australia; BMWF, Austria; ANAS, Azerbaijan; SSTC, Belarus; CNPq and
FAPESP, Brazil; NSERC, NRC and CFI, Canada; CERN; CONICYT, Chile; CAS,
MOST and NSFC, China; COLCIENCIAS, Colombia; MSMT CR, MPO CR and VSC
CR, Czech Republic; DNRF, DNSRC and Lundbeck Foundation, Denmark;
EPLANET and ERC, European Union; IN2P3-CNRS, CEA-DSM/IRFU, France;
GNAS, Georgia; BMBF, DFG, HGF, MPG and AvH Foundation, Germany; GSRT,
Greece; ISF, MINERVA, GIF, DIP and Benoziyo Center, Israel; INFN,
Italy; MEXT and JSPS, Japan; CNRST, Morocco; FOM and NWO, Netherlands;
RCN, Norway; MNiSW, Poland; GRICES and FCT, Portugal; MERYS (MECTS),
Romania; MES of Russia and ROSATOM, Russian Federation; JINR; MSTD,
Serbia; MSSR, Slovakia; ARRS and MVZT, Slovenia; DST/NRF, South
Africa; MICINN, Spain; SRC and Wallenberg Foundation, Sweden; SER,
SNSF and Cantons of Bern and Geneva, Switzerland; NSC, Taiwan; TAEK,
Turkey; STFC, the Royal Society and Leverhulme Trust, United Kingdom;
DOE and NSF, United States of America.

The crucial computing support from all WLCG partners is acknowledged
gratefully, in particular from CERN and the ATLAS Tier-1 facilities at
TRIUMF (Canada), NDGF (Denmark, Norway, Sweden), CC-IN2P3 (France),
KIT/GridKA (Germany), INFN-CNAF (Italy), NL-T1 (Netherlands), PIC
(Spain), ASGC (Taiwan), RAL (UK) and BNL (USA) and in the Tier-2
facilities worldwide.

\bibliographystyle{elsarticle-num}
\bibliography{h2bbpaper}
  
\clearpage

\begin{flushleft}
{\Large The ATLAS Collaboration}

\bigskip

G.~Aad$^{\rm 48}$,
B.~Abbott$^{\rm 111}$,
J.~Abdallah$^{\rm 11}$,
S.~Abdel~Khalek$^{\rm 115}$,
A.A.~Abdelalim$^{\rm 49}$,
O.~Abdinov$^{\rm 10}$,
B.~Abi$^{\rm 112}$,
M.~Abolins$^{\rm 88}$,
O.S.~AbouZeid$^{\rm 158}$,
H.~Abramowicz$^{\rm 153}$
% ...
\bigskip

% ...
$^{10}$ Institute of Physics, Azerbaijan Academy of Sciences, Baku, Azerbaijan\\
$^{11}$ Institut de F\'isica d'Altes Energies and Departament de F\'isica de la Universitat Aut\`onoma  de Barcelona and ICREA, Barcelona, Spain\\
% ...
$^{48}$ Fakult\"{a}t f\"{u}r Mathematik und Physik, Albert-Ludwigs-Universit\"{a}t, Freiburg i.Br., Germany\\
$^{49}$ Section de Physique, Universit\'e de Gen\`eve, Geneva, Switzerland\\
% ...
$^{88}$ Department of Physics and Astronomy, Michigan State University, East Lansing MI, United States of America\\
% ...
$^{111}$ Homer L. Dodge Department of Physics and Astronomy, University of Oklahoma, Norman OK, United States of America\\
$^{112}$ Department of Physics, Oklahoma State University, Stillwater OK, United States of America\\
% ...
$^{115}$ LAL, Universit\'e Paris-Sud and CNRS/IN2P3, Orsay, France\\
% ...
$^{153}$ Raymond and Beverly Sackler School of Physics and Astronomy, Tel Aviv University, Tel Aviv, Israel\\
% ...
$^{158}$ Department of Physics, University of Toronto, Toronto ON, Canada\\
% ...
\end{flushleft}


\end{document}